\documentclass[11pt]{article}

\usepackage{enotez}
\usepackage[utf8]{inputenc}
\usepackage[bottom = 1.5in, top = 1.5in, total={6.6in, 9in}]{geometry}
\usepackage{libertine}
\usepackage{graphicx}
\usepackage{epigraph}
\usepackage{comment}
\usepackage[table]{xcolor}

\usepackage{amsfonts,amsmath,amssymb,amsthm,mathrsfs,url,hyperref, graphicx, pxfonts, xcolor}

\usepackage{physics}
\usepackage{microtype}
 \usepackage[labelfont=bf]{caption}
\usepackage{multirow}
\usepackage{amsmath}
\usepackage{amssymb}
\usepackage{cleveref}
\usepackage{ragged2e}
\usepackage{natbib}

\newcommand{\x}[1]{{\color{black}#1}}
\newcommand{\y}[1]{{\color{black}#1}}
\newcommand{\z}[1]{{\color{black}#1}}

\title{\vspace{-5em} Putting Pressure Under Pressure: \\ On the Status of Classical Pressure in Special Relativity}

\author{Eugene Y. S. Chua\thanks{School of Humanities, Nanyang Technological University Singapore. Email: eugene.chuays@ntu.edu.sg}.}
\date{Accepted at \textit{Synthese}. Preprint of 30 January 2026.}
\begin{document}
\maketitle

\vspace{-2em}
\begin{abstract}
  \noindent \normalsize Much of the century-old debate surrounding the status of thermodynamics in relativity has centered on the search for a suitably relativistic temperature; recent works by \citet{chua_t_2023} and \citet{doi:10.1086/737745} have suggested that the classical temperature concept -- consilient as it is in classical settings -- `falls apart' in relativity. However, these discussions typically assume an unproblematic Lorentz transformation for -- specifically, the Lorentz \textit{invariance} of -- the pressure concept. Here I argue that, just like the classical temperature, the classical concept of pressure breaks down in relativistic settings. I discuss how this might suggest a new thermodynamic limit -- a $\textbf{u} \to 0$ limit -- without which an unambiguous thermodynamic description of systems doesn't emerge.
\end{abstract}

\begingroup
\tableofcontents
\endgroup

\RaggedRight
\setlength\parindent{2em}

\section{The question of relativistic thermodynamics}

The empirical success of classical thermodynamics has often supported a belief in its ``absolute authority" (\cite{uffink_bluff_2001}), giving the impression that thermodynamics is ``about almost everything" (\cite{atkins_four_2007}).\footnote{By classical thermodynamics, I mean what's usually called the Minus-First, Zeroth, First, Second, and Third laws. I assume this to be conceptually distinct from, though intricately connected to, statistical mechanics.} Likewise, Einstein's principle of relativity -- the idea that the quantities and laws of our physical theories should be invariant under Lorentz transformations -- is often touted as a principle to which we can ``ascribe general and absolute accuracy" (\cite{planck_dynamics_1908}). Thus emerged the question of \textit{relativistic thermodynamics}, that is, whether (and how) classical thermodynamics fits with the principle of relativity.\footnote{See e.g. \citet{einstein_relativity_1907} or \citet{planck_dynamics_1908}.}  Specifically, can we find Lorentz transformations of the laws and quantities of thermodynamics which preserve or `naturally' extend the physical meaning of these laws and quantities in the classical domain?\footnote{Relativistic thermodynamics also raises new philosophical worries about the limits of thermodynamics' applicability to relativistic settings, and hence to philosophical debates, e.g., concerning a global thermodynamic arrow of time \citep{albert_time_2000, loewer_mentaculus_2019, chen_quantum_2021} or whether black holes are thermodynamic objects \citep{dougherty_black_2016, wallace_case_2018}.}

\z{I focus on the limits of this project for relativizing the classical concept of \textit{pressure} as it appears in classical thermodynamics (henceforth simply $p_{\text{classical}}$).} I'll consider four perspectives (\S3) which together defined a unique $p_{\text{classical}}$ in the classical setting -- via hydrostatics, via the fundamental relation, via empirical equations of state, and via continuum mechanics -- and argue that their relativistic counterparts no longer jointly define a unique pressure concept in the relativistic domain (\S4). There is no unique way to extend the concept of pressure into the relativistic domain; it is only unique in the special case where systems are \textit{not} moving with respect to observers. This result sheds new light on the limits of thermodynamics and the usual thermodynamic limit (\S5).

\section{Relativizing thermodynamics: `T falls apart'}

To render thermodynamics compatible with the principle of relativity is to (at least) find a set of Lorentz transformations for the laws and quantities of classical thermodynamics -- by which we can relate thermodynamic quantities of systems in relative motion to some stationary observer (`in the moving frame' of the system) to the quantities of such systems as it appears to observers co-moving with the system (`in the rest frame' or `co-moving frame' of the system).\footnote{See \citet{liu_einstein_1992, liu_is_1994}.} These Lorentz transformations should either preserve the physical meaning of the original physical concepts and laws, or provide natural extensions to them -- it's one thing to find some set of transformations for some quantity, and another to assert that the quantity preserves its physical meaning under such transformations.

Lorentz transformations allow us to relate the time and spatial coordinates between two inertial frames in relative motion (and hence quantities depending on such coordinates). Consider an observer $ O' $  with time and position coordinates $(t', x', y', z')$ , moving along the positive $x$-axis relative to another observer $ O $  with coordinates $ (t, x, y, z)$ . If $O'$ is moving in the $x$-direction at a constant velocity $\textbf{u}$ (where $\textbf{u} = \textbf{v}/c$ and $\textbf{v}$ is the relative velocity of $O'$ from $O$), then $O$  can describe $O'$ 's coordinates using the Lorentz transformations:
\begin{equation}
    t' = \gamma (t - ux),  \; \; \; \; x' = \gamma (x - ut),  \; \; \; \;
    y' = y, \;  \; \; \; z' = z
\end{equation}
where $\gamma = \frac{1}{\sqrt{1 - u^2}}$ is the Lorentz factor, $u$ is the magnitude of $\textbf{u}$, and $c$ is set to 1 for simplicity for the rest of this paper (so that $0 \leq u \leq 1$). 

The hope for (special-)relativistic thermodynamics is to find unique Lorentz transformations for thermodynamic quantities like temperature, pressure, or volume. To that end, Planck and Einstein proposed a set of Lorentz transformations for most thermodynamic quantities.\footnote{See Liu (1994) and references therein.} Their proposed Lorentz transformations for the volume element $dV$, pressure $p$, and entropy $S$ (and energy, to be discussed in \S4.2 and \S4.4) are:

\begin{equation}
    dV' = \frac{dV}{\gamma},  \; \; \; \; p' = p,  \; \; \; \; S' = S
\end{equation}
The above remains largely uncontroversial (though we'll return to pressure shortly). However, the Lorentz transformation for \textit{temperature} turned out to be extremely equivocal,\footnote{See \citet{liu_einstein_1992, liu_is_1994, chua_t_2023} and references therein.} which led Einstein to remark that, when it comes to relativizing thermodynamics,
\begin{quote}
    there is actually no compelling method in the sense that one view would simply be `correct' and another `false.' One can only try to undertake the transition \textit{as naturally as possible}. (\cite[p. 200, emphasis mine]{liu_einstein_1992})
\end{quote}
\citet{chua_t_2023} provides one way of understanding this notion of `natural'-ness via \textit{consilience}. He suggests that
\begin{quote}
    we understand Einstein’s notion of `natural'-ness as follows: There’s strong consilience between classical counterparts of these procedures in determining [the classical temperature]’s physical meaning, in the operational sense that the temperature established by each procedure agrees with other procedures. Contrariwise, their relativistic counterparts demonstrate no such consilience: different procedures predict starkly different behaviors for relativistic temperature. `Natural' procedures in [classical thermodynamics] do not appear `natural' in relativistic settings. (Chua 2023, 1308)
\end{quote}
As \citet{chua_t_2023} shows, (at least) four procedures jointly defined the temperature in the classical setting: the Carnot cycle, thermometers, kinetic theory, and black-body radiation. Crucially, these four procedures all agree on the same concept of the classical temperature, each capturing different facets of its physical meaning. They are \textit{consilient} in defining the classical temperature. 

\y{Chua's consilience dovetails with Weyl's \textit{concordance}, with some subtle differences. It's worth briefly comparing the two, in order to elucidate the notion of consilience.\footnote{I thank an anonymous reviewer for drawing my attention to this connection between consilience and concordance.} Weyl’s focus is on the convergence of ``determinations" -- numerical inferences from empirical data and the established laws of physical theory -- about the values of physical quantities such as the electronic charge $e$. If a concept is concordant, then every determination of that concept’s value agrees:}
\begin{quote}
    \y{\textit{Every such determination has to yield the same result.} Thus all determinations of the electronic charge $e$, that follow from observations in combination with the laws established by physical theory, lead to the same value of $e$ (within the accuracy of the observations). \citep[121]{weyl1949philosophy}}
\end{quote}
\y{For Weyl, a properly physical quantity must be concordant in this sense, in order to ``bring the theory in contact with experience" \citep[122]{weyl1949philosophy}. Chua's consilience agrees that convergence of a certain sort is important for identifying whether a concept makes contact with -- and tracks -- reality. However, consilience is at once stronger and weaker. 

On the one hand, consilience appears stronger (and less strictly positivistic) than Weyl’s concordance, because it demands more than mere numerical agreement across measurement procedures. Specifically, consilience requires that various procedures share a deeper semantic unity, arising from their convergence in terms of theoretical roles. For example, in the case of temperature, the relevant procedures -- Carnot cycles, thermometers, kinetic theory, and blackbody radiation -- not only produce numerically consistent results but also jointly define what temperature \emph{means} in the classical domain. Thus, consilience involves not merely convergence in measurement outcomes, but convergence in the explanatory and conceptual roles that jointly define a unified physical quantity.

On the other hand, consilience appears to be weaker, in that it can be \textit{domain-specific}. Weyl's discussion of concordance does not appear to be domain-specific: a quantity either is or is not genuinely physical, and, if it is, then \textit{all} determinations must match. But Chua holds that classical temperature can be perfectly consilient---and thus perfectly real---only in ordinary, non-relativistic contexts involving Carnot cycles, thermometers, kinetic theory, and black-body radiation. In relativistic contexts, the relativistic generalizations of these classical procedures generalize in divergent, inconsistent ways: different approaches yield different relativistic transformations for the temperature, \(T' = T,\, T' = \gamma T,\, T' = \frac{1}{\gamma}T\), and can even be arbitrary. In short, there is no single relativistic temperature that carries the same convergent meaning -- the same consilience -- as in the classical domain.\footnote{\z{Finally, there is a rather deep question here, concerning what to say about a concept when the breakdown of consilience occurs. In my view, we can say one of two things: (1) \textit{the} concept of temperature/pressure breaks down simpliciter and only lesser concepts which are not temperature/pressure (because they no longer play all the roles we expect of the classical concept) survive the generalization. Or (2) there really were many different sub-concepts of temperature or pressure that happened to align in the equilibrium regime and not elsewhere. This is an interesting question in philosophy of language, in part about whether and when we are using one concept as opposed to indeterminately using many. In this paper, I simply try to establish the prior claim, \textit{that} the consilience of $p_{\text{classical}}$ breaks down. The further question, of arbitrating between (1) and (2), will take us too far afield into deep questions in philosophy of language, and I leave it for future work.}} Consequently, classical temperature ``falls apart” in relativity; a concept's consilience can break down in the relativistic domain even as the concept remains robust (and tracks reality) within the classical domain.\footnote{\y{In this sense, I would likely agree with \citet[16]{bridgman1927logic}'s remark that a mere numerical match -- such as that between tactile and optical length -- \textit{need not} imply that the two concepts are physically the same. Determining whether they are indeed physically the same is a more subtle question. As an operationalist, Bridgman holds that (as in the case of tactile vs. optical length) ``if we have more than one set of operations, we have more than one concept, and strictly there should be a separate name to correspond to each different set of operations” (p. 11). Because I am not an operationalist, I am open to the idea that distinct theoretical procedures -- including but not limited to measurement operations -- can define a single consilient concept. In temperature’s case, physical practice typically does not separate “kinetic temperature” from “Carnot temperature,” or from “blackbody radiation temperature”: instead, practitioners treat them as one classical concept. Yet Bridgman’s operationalism would generate many separate concepts even in this case where one ``natural" concept, such as classical temperature, can unify them. Nonetheless, my analysis aligns with the spirit of Bridgman’s caution that “we must always be prepared some day to find that an increase in experimental accuracy may show that the two different sets of operations which give the same results in the more ordinary part of the domain of experience, lead to measurably different results in the more unfamiliar parts of the domain” \citep[24]{bridgman1927logic}. As argued above (and will be argued later), consilience in one domain does not guarantee consilience in another. I thank an anonymous reviewer for suggesting that I further develop these connections between Bridgman's discussion and consilience.}}}

These problems with relativistic thermodynamics have so far only been discussed in the context of temperature. Importantly, Einstein, Planck, and others took themselves to have definitively obtained the Lorentz transformation of pressure $p$:
\begin{equation}\label{pinvariance}
    p' = p
\end{equation}
That is, $p$ is a Lorentz-invariant quantity. Discussions about the special relativistic temperature have largely proceeded on this understanding, seen in e.g. \citet{landsberg_a_1967, liu_is_1994, chua_t_2023}. For instance, \citet[29]{landsberg_a_1967} labels \eqref{pinvariance} as ``generally valid and accepted". However, as we'll see, the problem with relativistic thermodynamics is not just a problem with temperature, something only a few e.g. \citet{farias_what_2017, sutcliffe_lorentz_1965, balescu_relativistic_1968} have recognized.

\section{The classical thermodynamic pressure}

\z{The goal for the remainder of this paper is precisely to work out what happens when $p_{\text{classical}}$ is generalized to the relativistic domain. In particular, what happens to it under a Lorentz boost? For concreteness, by $p_{\text{classical}}$, I mean the unique equilibrium quantity featuring in non-relativistic equilibrium thermodynamic regimes. In this regime, $p_{\text{classical}}$ is jointly defined in terms of four roles it plays, which I will explain in the remainder of this section:
\begin{itemize}
    \item \textbf{Hydrostatic role}: it is an isotropic quantity defined in terms of balanced forces acting over some surface area.
    \item \textbf{Thermodynamic role}: it features as an intensive parameter in the fundamental relation.
    \item \textbf{Empirical role}: it features as a parameter in a variety of empirical equations of state for specific kinds of systems. 
    \item \textbf{Continuum mechanical role}: it is defined in the local rest frame of static or ideal fluids as the isotropic part of the stress tensor.
\end{itemize}
The question, then, is what happens when this concept, understood as such, is generalized to the relativistic domain. As with the relativistic temperature, the question is what happens to this concept under Lorentz boosts.}

\x{Naively, }one might think that the Lorentz invariance of pressure follows from the fact that $p_{\text{classical}}$ is scalar. That $p_{\text{classical}}$ is scalar is certainly true in classical physics, though \textit{why} it is true is not quite straightforward.

\x{Furthermore, it helps to distinguish two senses of a `scalar' quantity here.\footnote{I thank an anonymous referee for suggesting this distinction.} One is to define a scalar simply as a \textit{Lorentz-invariant quantity}. If $p_{\text{classical}}$ is indeed scalar in this sense, then it trivially follows that it is Lorentz-invariant. Of course, this is precisely what I'll call into question later.

Rather, $p_{\text{classical}}$ is a `scalar' in another sense: it is a real-valued quantity, acting on vectors via scalar multiplication. $p_{\text{classical}}$ is indeed scalar in this sense, and we can define a scalar field in terms of the value of $p_{\text{classical}}$ at each point of a system. The classical temperature is also scalar in this sense, and I have already mentioned that the question of its Lorentz-invariance is far from closed. As I'll show, there's likewise no guarantee that $p_{\text{classical}}$ is Lorentz-invariant just because it is scalar in \textit{this} sense.}

To see how $p_{\text{classical}}$ breaks down under pressure, I'll first show how it can be understood uniquely in the classical setting, through the consilience of four different perspectives in defining $p_{\text{classical}}$.\footnote{Crucially, I am not saying that these perspectives are \textit{independent} -- they cannot be independent if they are consilient, because they \textit{build on each other}. But they are not conceptually identical, and this lack of conceptual identity is what allows us to see the breakdown of consilience in the relativistic regime.}

\subsection{Pressure in hydrostatics}

\z{Let us first consider the \emph{hydrostatic} role that $p_{\text{classical}}$ plays: in equilibrium fluids, $p_{\text{classical}}$ is an \emph{isotropic scalar} magnitude of balanced normal forces per unit area. This fixes one component of the classical concept that we will compare against its relativistic counterparts in \S4.} 

$p_{\text{classical}}$ is often associated with forces over unit area, but forces are always directed. In modern terms, forces are vectorial, and can furthermore vary over different directions in space, i.e. they can be anisotropic. Yet $p_{\text{classical}}$ is both scalar (in the second sense from the foregoing) and \textit{isotropic}: \x{the scalar field defined by $p_{\text{classical}}$ over some system is not supposed to vary differently with any particular direction in space.} An immediate puzzle is how one gets a scalar isotropic quantity from a vectorial and generally anisotropic quantity. Indeed, the idea that pressure might not depend on direction at all is fairly new. To my knowledge, this understanding was first developed in terms of the pressure of fluids in equilibrium, in the study of \textit{hydrostatics}.\footnote{See \citet{chalmers_how_2018, chalmers_one_2017} for a history of pressure and hydrostatics.} It is in this context that the isotropy of pressure became \textit{de rigueur}. Today, we can readily find claims such as \citet[pp. 13--14]{massey_mechanics_2006}'s emphatic statement that ``to say that pressure acts in any direction, or even in all directions, is meaningless; pressure is a scalar quantity." The first fully modern understanding of pressure as a scalar isotropic quantity, though, is found in \citet[Book II, \S5]{newton_principia_1687},\footnote{See \citet{chalmers_how_2018}.} which relied on hydrostatic equilibrium and the idea of \textit{balanced forces}, and explicitly defined pressure in terms of \textit{forces over area}.

\begin{figure}[ht]
    \centering
    \includegraphics[scale = 0.25]{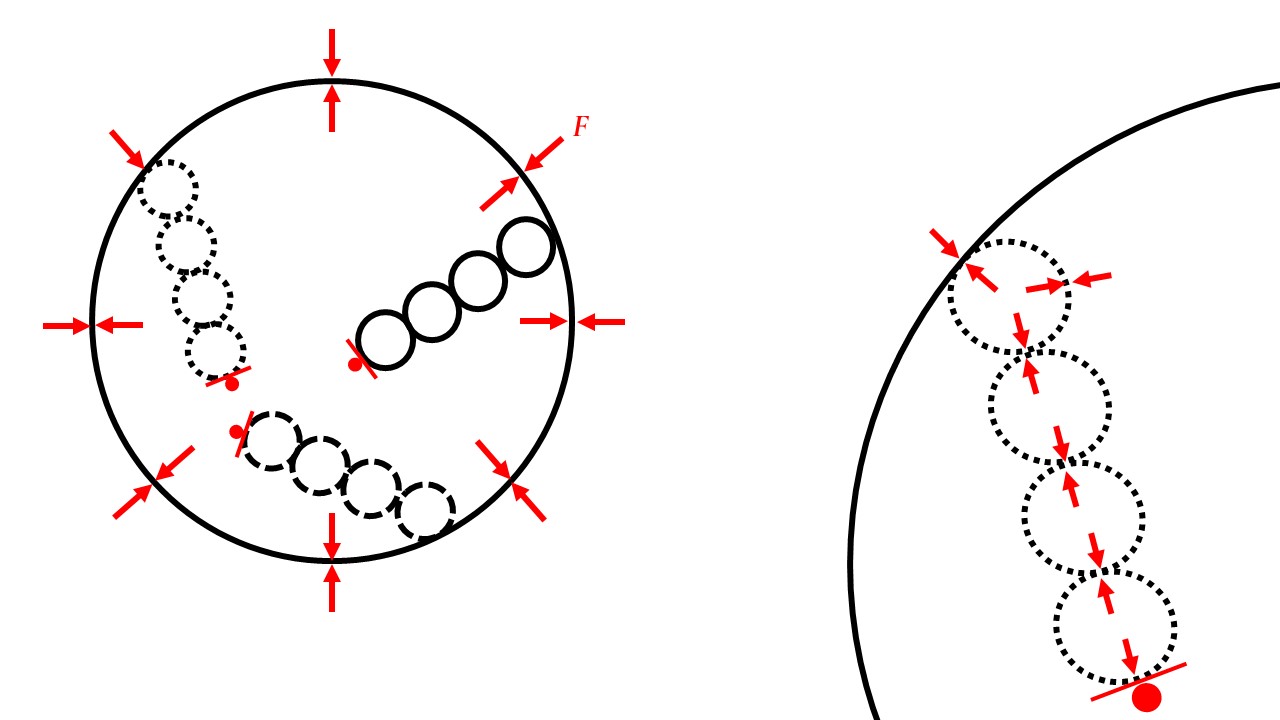}
    \caption{Newton's argument for the isotropy of pressure, illustrated.}
    \label{fig:1}
\end{figure}

Newton famously argued for the isotropy of pressure, through which we can see \textit{why} pressure is a scalar.\footnote{I largely follow \citet{chalmers_how_2018}'s presentation.} Consider a fluid in a sphere,\footnote{A fluid is ``any body whose parts yield to any force applied to it and yielding are moved easily with respect to one another." \citep[Book II, \S5]{newton_principia_1687} } pressed on all points of its outer surface by incoming vectorial forces $\textbf{F}$ per unit area with uniform magnitudes $|\textbf{F}|$. (See fig. \ref{fig:1}.) Newton assumed that the fluid is incompressible,\footnote{They do not change in volume or density under changes in pressure.}  and so, Newton reasons, all of the parts of the fluid must be \textit{static}: there is no distortion and motion -- no net force -- in each part of the fluid. This implies a balance of forces at any point on the surface: for every unit area on the sphere's surface, there is a force $-\textbf{F}$ with equal magnitude but opposite direction. 

\y{To show the isotropy of pressure throughout the fluid is to show that these balanced forces, acting on any arbitrary plane at \textit{any arbitrary point} \textit{within} the sphere (the red lines and dots in fig. \ref{fig:1}), have the same magnitudes everywhere. Given any arbitrary point in the fluid, imagine a chain of small, adjacent sub-spheres leading from that point to the fluid’s outer surface, where the incoming force $\textbf{F}$ first acts. By Newton’s third law, each sub-sphere exerts and experiences equal and opposite forces with its neighbors; because the fluid is static, all points of each sub-sphere’s surface must be under a uniform force of magnitude $|\textbf{F}|$. This uniform balance extends inward step-by-step, so that all sub-spheres must experience forces per unit area with the same magnitude, \textit{balanced} with forces of equal magnitude but opposite direction, in order for the fluid to remain static. Hence, the \textit{magnitude} of these balanced forces is identical everywhere, and we define this scalar as the hydrostatic pressure $p$.}

From this perspective, pressure is \textit{neither} conceptually identical to the directed forces acting on the sphere \textit{nor} the directed balancing forces, even though it cannot be defined without them. It is the \textit{magnitude} of the \textit{balanced forces} per unit area everywhere in a static fluid \citep[p. 174]{chalmers_how_2018}. 

In short, the hydrostatic pressure's isotropy and scalar nature depend on two conditions on the forces per unit area: their magnitudes are constant in all directions, and each of these forces must be balanced with equal and opposite forces. Thus, while the hydrostatic pressure is sometimes said to be \textit{defined as forces over area}, it is not strictly correct. The hydrostatic pressure is not equivalent to \textit{forces} \textit{per se}; it is equivalent to the \textit{magnitude} of isotropic balanced forces, i.e. \textit{in hydrostatic equilibrium}, which allows us to define it directly as a single, scalar, quantity. 

\z{As an aside, it is worth noting that Newton, in \textit{De Gravitatione}, took fluids to be strictly continuous:\footnote{As \citet{chalmers_how_2018} notes, this is not too far off from the foundations of contemporary hydrostatics. Newton's argument -- and the assumption that fluids are strictly continuous --survives in contemporary hydrostatics: ``Newton’s proofs were dependent on the strict continuity assumption insofar as they involved spheres of liquid touching planes at points. The same continuity remains in the modern formulation of hydrostatics set in place by Euler. It is implicit in the deployment of the differential equations that replaced Newton’s geometrical diagrams."} ``Thus I imagine that a fluid does not consist of hard particles, but that it is of such a kind that it has no small portion or particle which is not likewise fluid." (\citet[151]{Newton1962USP}, see also \citet{chalmers_how_2018} for the tension between this assumption of continuity and Newton's broader corpuscular commitments.) Nowadays we think that fluids are composed of particles so that at some scales the argument might fail. However, this is no issue for our purposes, which is to specify features of the hydrostatic role of pressure in regimes of thermodynamic equilibrium: insofar as we are in regimes where systems can be treated as fluids on which forces are impressed, the hydrostatic argument for isotropy continues to hold (and is fairly standard textbook fare). In any case, my point here is \textit{not} that Newton's entire account of hydrostatics -- ontology and all -- is correct. Rather, it is simply a modest point that particular features of the pressure, that it is isotropic and can be understood in terms of balanced forces over unit areas, stem from Newton's account and remain in use today as a defining trait of $p_{\text{classical}}$.} 

\z{While this feature -- that the hydrostatic pressure is isotropic -- was first established through Newton's argument which holds only for incompressible fluids, the result holds in more general cases as well.  Notably, \citet{euler_principes_1757} subsequently generalized Newton's result to \textit{all} fluids in equilibrium  -- incompressible and compressible -- which included gases like air. Additionally, this understanding of $p_{\text{classical}}$ as an isotropic magnitude of balanced forces over area for fluids in equilibrium -- now including gases -- allowed later physicists to connect thermodynamics to statistical mechanics via the kinetic theory of heat and gases. In the case of the ideal gas, it can be easily proven that the pressure of the ideal gas is proportional to the mean (translational) kinetic energy of the gas particles and the number of particles per unit volume.\footnote{See, for instance, \citet[ch. 21]{halliday_fundamentals_2007} for details.} This is the classical result (with $\Bar{v}^2$ the gas particles' mean squared velocity, $N$ the particle number, and $V$ the ideal gas's total volume:\footnote{Early versions of this result can be found in e.g. \citet{Clausius1857-Heat-PhilMag}.}
\begin{equation}\label{pressure-energy}
    p = \frac{2}{3}\frac{N}{V}\frac{1}{2}m\Bar{v}^2
\end{equation}
This result, in particular, enabled the connection of our understanding of $p_{\text{classical}}$ directly to mechanical quantities (in terms of momenta, kinetic energy, and forces). The understanding of $p_{\text{classical}}$ in terms of the magnitude of balanced forces over area plays a crucial role in bridging the two since forces can be defined in terms of changes in momenta. 
Crucially, this result requires assuming the very same isotropy of pressure established above in Newton's proof which still holds in fluid regimes, but can also be justified by appealing to the random motion of particles which has no preferred direction \citep[ch. 21]{halliday_fundamentals_2007}. 

To sum up the foregoing discussion, one prominent understanding of pressure is in terms of hydrostatics, as an isotropic scalar quantity as a result of balanced forces over areas (we see in \S4.1 that Einstein has such a picture in mind when he proves the isotropy of pressure). However, this isotropy is crucially dependent on background assumptions: specifically, pressure can be understood as an isotropic scalar quantity \textit{if} the fluid is (or becomes) static, and the system equilibrates with its environment (i.e. the forces acting on the fluid are balanced against the forces acting on the environment). In \S4.1 we revisit exactly how this hydrostatic role generalizes under Lorentz boosts.}

\subsection{Pressure in the fundamental relation}

While the hydrostatic understanding of pressure as the magnitude of balanced forces over area is important, it does not \textit{exhaust} the understanding of $p_{\text{classical}}$. Let us now turn from the mechanical picture of pressure to the \textit{classical thermodynamic framework}. In this framework, at least classically, the isotropy of $p_{\text{classical}}$ is retained via its role as an intensive state variable (hence its consilience with hydrostatics and mechanics since intensive variables do not scale with size -- they remain isotropic). However, this thermodynamic understanding of pressure is much richer -- and different -- as it connects $p_{\text{classical}}$ to the other thermodynamic state variables.\footnote{A state variable depends only on the current thermodynamic equilibrium state of the system, independent of the path taken to reach it.}    

Following Callen,\footnote{A similar approach is found in e.g. \citet[pp. 31 -- 35]{reichl_modern_1980}.} one standard way of viewing thermodynamics from a fundamental perspective involves understanding the system's conserved internal energy $U$ (or, in an alternative but equivalent representation, the conserved entropy $S$) as a function of all the thermodynamically relevant and independent extensive parameters (e.g. entropy $S$, volume $V$, mole numbers $N_i$) which exhaustively defines any possible thermodynamic state. \citet[p. 15]{callen_thermodynamics_1991} sees them as the drivers of change in thermodynamics: by manipulations of the `walls' or constraints of the system, ``the extensive parameters of the system are altered and processes are initiated". Each extensive parameter, then, represents an independent way of manipulating the system through thermodynamic means, and hence also defines a class of thermodynamic processes through which one only manipulates them and not other extensive parameters.

Writing down $U$ (or $S$) in terms of the other independent extensive parameters, we obtain the \textit{fundamental equation}, or, equivalently, the \textit{fundamental relation}. In Callen's terms:
\begin{quote} 
   The information contained in a fundamental relation is \textit{all-inclusive} -- it is equivalent to \textit{all conceivable numerical data, to all charts, and to all imaginable types of descriptions of thermodynamic properties}. If the fundamental relation of a system is known, \textit{every thermodynamic attribute is completely and precisely determined}. (\cite[p. 28]{callen_thermodynamics_1991}, emphasis mine)
\end{quote}
The fundamental relation is thus the theoretical bedrock of thermodynamics: knowing it amounts to knowing all possible thermodynamic behavior of the system. The usual form of the fundamental equation (in the $U$-representation) is:
\begin{equation}
    U = U(S, V, N_1, ... N_r)
\end{equation}
That is, $U$ is a function of some set of independent extensive parameters: each parameter can be changed, in principle, independent of the others (e.g. we can quasi-statically manipulate volume while preserving entropy). Taking the total differential of $U$ in terms of these parameters:
\begin{equation}
    dU = \left(\frac{\partial U}{\partial S}\right)\Bigg|_{V, N_1,...,N_r} dS + \left(\frac{\partial U}{\partial V}\right)\Bigg|_{S, N_1,...,N_r} dV + \sum_{j = 1}^{r} \left(\frac{\partial U}{\partial N_j}\right)\Bigg|_{S, V, ... N_r} dN_j 
\end{equation}
This looks just like the first law of thermodynamics: 
\begin{equation}
    dU = dQ + dW
\end{equation}
from which we can define the partial derivatives in terms of familiar results. For our purposes, we can use the familiar relation for compressional work via changes in volume, $dW = -pdV$, to identify:
\begin{equation}\label{pVeq}
\left(\frac{\partial U}{\partial V}\right)\Bigg|_{S, N_1,...,N_r} := -p 
\end{equation}
Here, pressure $ p $ is \textit{not defined in terms of the magnitude of balanced forces over area}. Rather, it is defined as the partial derivative of internal energy $U$ with respect to volume $V$, holding $S$ and $N_j$ (plus any other extensive variables) constant.  Concretely, consider a scenario where a system undergoes a change in volume, but maintains constant entropy and chemical composition. The associated total change in internal energy $dU$ manifests as work $dW$, which is linked to pressure via $dW = -pdV$. Therefore, in this framework, pressure tracks energy variations tied to changes in the system's volume. This is related to but quite \textit{conceptually distinct from its role as a measure of the magnitude of balanced mechanical forces over area}.

In the classical framework, there is remarkable consilience between these two understandings of pressure. As Callen (pp. 49--51, emphasis mine) asserts: ``the pressure defined by [\eqref{pVeq}] \textit{agrees in every respect} with the pressure defined in mechanics." Just as the hydrostatic pressure tracks the magnitude of balanced forces per unit area across a system in hydrostatic equilibrium, the thermodynamic pressure is also balanced across different parts of a system in thermodynamic equilibrium (that is, a state in which all thermodynamic quantities are stationary over time).\footnote{\y{Consider two subsystems $S_1$ and $S_2$ in a closed composite system in thermodynamic equilibrium, separated by a movable adiabatic and impermeable wall. While $S$ and $N_j$ remain fixed, $U_1 + U_2 = \text{const.}$ and $V_1 + V_2 = \text{const.}$ imply $dU = dU_1 + dU_2 = 0$ and $dV_1 = -dV_2$. Applying the fundamental relation, we get:
\[
dU = \left(\frac{\partial U_1}{\partial V_1}\right)_{S_1,N_1} dV_1 + \left(\frac{\partial U_2}{\partial V_2}\right)_{S_2,N_2} dV_2 = (p_2 - p_1) \; dV_1 = 0,
\]
so $p_1 = p_2$. The subsystems equilibrate by adjusting volumes until their pressures match.}} In this sense, the two perspectives -- thermodynamic and hydrostatic -- seems to be different ways, classically, of understanding the one selfsame concept $p_{\text{classical}}$. However, in \S4.2 we see that this consilience breaks down: the thermodynamic role departs from the hydrostatic one upon considerations about Lorentz boosts.

\subsection{Pressure in equations of state}

While the fundamental relation is the \textit{theoretical} bedrock of thermodynamics, its specific form for particular systems is often not available to us, and hence its empirical import is limited: ``In thermodynamic theory... we accept the existence of the fundamental equations, but we do not assume explicit forms for them, and we therefore do not obtain explicit answers." (\cite[p. 44]{callen_thermodynamics_1991}) In practice, much of the work in thermodynamics involves finding the \textit{equations of state} (EOS) for specific systems, from which we attempt to recover the fundamental relation for said system. EOS provide the \textit{empirical} content of thermodynamics, connecting the theoretical structure of the fundamental relation to concrete systems in question, be it terrestrial materials, or the interiors of planets and stars.\footnote{See e.g. \citet{anderson_equations_1995}.} Such EOS provides empirical meaning for $p_{\text{classical}}$, by telling us how pressure behaves concretely for specific systems: they \textit{characterize} the system's specific empirical relationship between pressure and other thermodynamic state variables which \textit{could not be discovered via the theoretical framework alone}. Crucially, the fundamental relation alone is too unconstrained to tell us about these specific empirical relations; they must be discovered empirically and only hold for specific types of systems. For instance, while the well-known Ideal Gas Law
\begin{equation}\label{idealgaslaw}
    pV = Nk_BT
\end{equation}
predicts that $p_{\text{classical}}$ will increase without bounds as $V$ decreases at constant $T$ and $N$, more realistic EOS tell us otherwise. Consider the Van der Waals equation of state:
\begin{equation}
    (p + \frac{a}{V_m^2})(V_m - b) = RT
\end{equation}
This modifies the ideal gas law by introducing two empirically discovered parameters: the volume occupied by gas molecules $b$, and $a$ which reflects intermolecular attractions. These adjustments are crucial for qualitatively explaining gas-liquid phase transitions,\footnote{It is well-known that the van der Waal's EOS is \textit{quantitatively} inaccurate with regards to experimental results.} which the ideal gas law fails to predict. More complicated EOS, e.g., the Birch-Murnaghan EOS, capture more physically interesting phenomena, which I won't get into here. 

What's important is that these considerations go beyond the scope of the abstract fundamental relation per se: different EOS are generally compatible with the fundamental relation. EOS thus provide a distinct perspective on thermodynamic quantities for specific systems, such as how $p_{\text{classical}}$ will actually behave in relation to other extensive and intensive parameters for specific types of systems, in consilience with the the abstract fundamental relation. Again, in the classical thermodynamic regime, there is robust consilience between the role that pressure plays in various EOS, and the hydrostatic and thermodynamic roles. However, we'll see in \S4.3 how this consilience breaks down. 

\subsection{Pressure in continuum mechanics}

The final perspective on $p_{\text{classical}}$ I'll discuss is from \textit{continuum mechanics}, in which a generalized notion of pressure is defined. The continuum mechanical picture provides a unified framework in which one approaches both fluids and solids. Crucially, from this perspective, pressure is \textit{not} taken to be isotropic, nor scalar, in general. However, in the domains where classical thermodynamics -- and $p_{\text{classical}}$ -- apply, the two perspectives will agree.

\begin{figure}
    \centering
    \includegraphics[scale = 0.25]{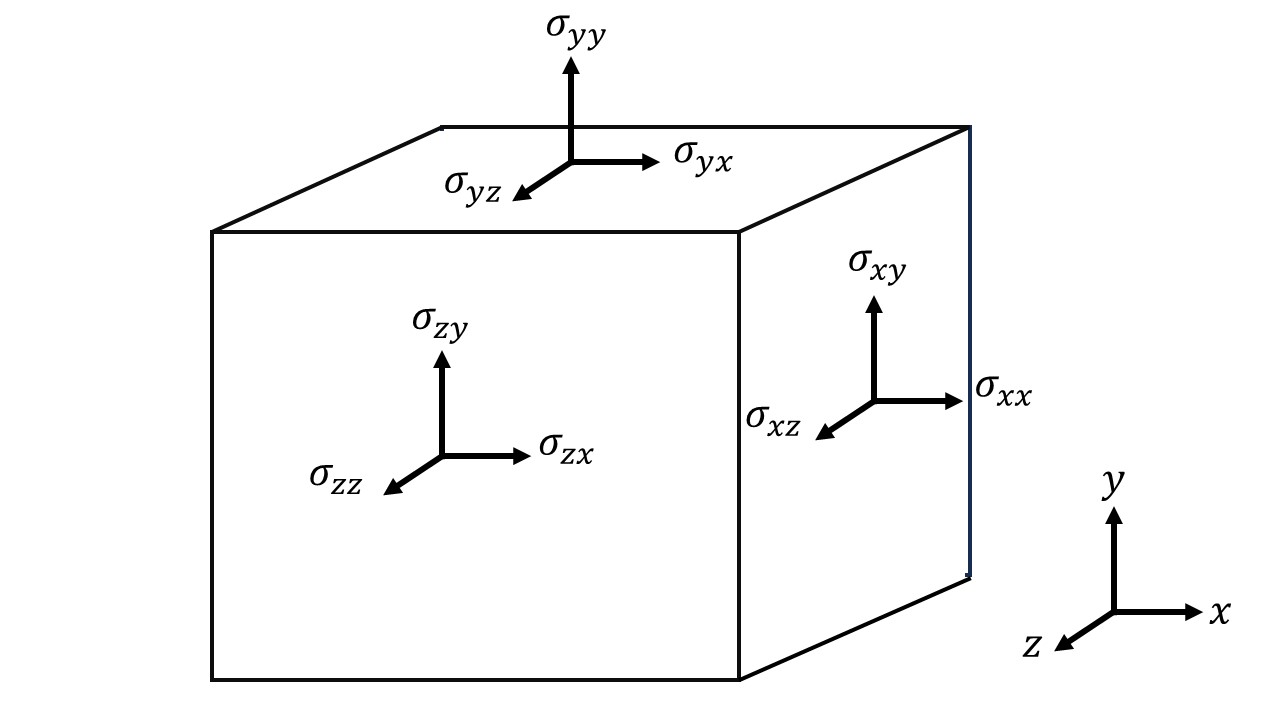}
    \caption{Adapted from \citet[p. 117]{tadmor_continuum_2012}.}
    \label{fig:2}
\end{figure}

Continuum mechanics formalizes the concept of \textit{stress}, that is, the types of internal forces which unit volume elements of bodies can experience, usually via the \textit{Cauchy stress tensor}. Crucially, and in consilience with the hydrostatic picture, $p_{\text{classical}}$ can be understood in terms of one such force. However, in this picture, the hydrostatic understanding of pressure is simply a special case. As \citet{cauchy_pression_1827} suggests, defined in terms of general stresses,\footnote{This translation is from \citet[Ch. 3]{truesdell_first_1992}.}
\begin{quote}
  ...the new pressure \textit{does not always remain perpendicular to the faces subject to it, nor is it the same in all directions at a given point}. (emphasis mine)
\end{quote}
For any volume element (see Fig. \ref{fig:2}), we can understand stresses to be acting either perpendicular, or normal to the unit surface area. Unlike the hydrostatic understanding of pressure, it is not assumed that \textit{the stress is isotropic} (that is, contra \citet{massey_mechanics_2006}, it is not `meaningless' to consider something like pressure acting differently in different directions). The magnitude of such forces in the various directions can be written down in terms of the aforementioned Cauchy stress tensor, denoted by $\sigma$.\footnote{See e.g. \citet[\S3]{truesdell_first_1992} or \citet[\S4.2.4--4.2.5]{tadmor_continuum_2012}.} Mathematically, it is represented as a $3\cross3$ matrix in three-dimensional space:
  \[\sigma = \begin{bmatrix} \sigma_{xx} & \sigma_{xy} & \sigma_{xz} \\ \sigma_{yx} & \sigma_{yy} & \sigma_{yz} \\ \sigma_{zx} & \sigma_{zy} & \sigma_{zz} \end{bmatrix} 
  \]
where $\sigma_{ij}$ represents the magnitude of the stress in the $i$\textsuperscript{th} direction acting on the plane normal to the $j$\textsuperscript{th} direction (see fig. \ref{fig:2}). Note that these are all \textit{internal} stresses -- exerted by the unit volume element (the opposing, balancing, forces, coming from an external system, are \textit{not included}. This becomes relevant in \S4.2.2 and \S4.4.)

There is a distinction between two sorts of stresses. On one hand, the diagonal elements ($\sigma_{xx}, \sigma_{yy}, \sigma_{zz}$) represent the magnitudes of \textit{normal stresses}, which are perpendicular to the respective planes. Crucially, normal stresses are just forces acting perpendicular to the respective planes (and the associated components their magnitudes).  On the other hand, the off-diagonal elements ($\sigma_{xy}, \sigma_{xz}, \sigma_{yx}, \sigma_{yz}, \sigma_{zx}, \sigma_{zy}$) represent \textit{shear stresses}, acting parallel to the plane. 

Now for the connection to hydrostatics. Hydrostatic systems can be understood as a special case of the Cauchy stress tensor under specific assumptions. In hydrostatics, we deal with fluids at rest, that is, \textit{static} fluids. In such a case, no part of the fluid is moving, and all forces acting on each part of the fluid act perpendicular to the part, i.e. they are normal stresses;  \citet[p. 54]{irgens_continuum_2008} \textit{defines} a fluid just to be ``a material that at rest only can transfer normal stresses on material surfaces". Put another way, static fluids experience no shear stresses, which means that all off-diagonal elements of the Cauchy stress tensor are zero:
   \begin{equation}\label{isotropyofp}
   \sigma_{xy} = \sigma_{xz} = \sigma_{yx} = \sigma_{yz} = \sigma_{zx} = \sigma_{zy} = 0 
   \end{equation}
Since the fluid is static, we again assume that the normal stresses are \textit{balanced} by incoming external forces (not represented in the stress tensor for the fluid!). Now, we can again evoke Newton's argument from $\S3.1$ and argue for the isotropy of these balanced normal stresses for static fluids. Hence, again, the diagonal elements of the stress tensor are equal; we can now \textit{define} the diagonal elements -- the magnitudes of these normal stresses -- as the hydrostatic pressure:
   \begin{equation} \sigma_{xx} = \sigma_{yy} = \sigma_{zz} = -p \end{equation}
Crucially, the magnitude $p$ can be defined as a scalar only because it is defined in terms of isotropic stresses, such that these stresses are balanced by external forces. 

Given these assumptions, the Cauchy stress tensor for hydrostatic systems simplifies to a form where all the diagonal elements are equal to each other and represent the hydrostatic pressure, and all off-diagonal elements are zero. The hydrostatic stress tensor looks like:
\begin{equation}
    \sigma = \begin{bmatrix} -p & 0 & 0 \\ 0 & -p & 0 \\ 0 & 0 & -p \end{bmatrix} 
\end{equation}
This recovers the hydrostatic picture in $\S3.1$, and hence supports a consilient understanding of $p_{\text{classical}}$. 

We can continue to use the same approach for \textit{nonstatic} fluids with variable velocity, if the fluid is an \textit{ideal} or \textit{perfect} fluid with \textit{zero viscosity}: equivalently, every volume element of the fluid experiences no shear stresses. Such a fluid experiences no friction within itself or with the container in which it flows. As \citet[p. 127]{tadmor_continuum_2012} notes, ``In an ideal nonviscous fluid there can be no shear stress. Hence, the Cauchy stress tensor is entirely hydrostatic, $\sigma_{ij} = -p\delta_{ij}$". There remains a well-defined (not necessarily constant) isotropic hydrostatic pressure for each fluid volume element. In these cases, we can still do thermodynamics and use $p_{\text{classical}}$ by e.g. assuming that each \textit{part} of the fluid is in \textit{local} thermodynamic equilibrium. (\cite{tadmor_continuum_2012}, p. 220). 


However, a branching point arises here. Beyond the above scenarios, there is no guarantee that the diagonal elements of $\sigma$ are identical. \textit{Generally}, the stress tensor's diagonal terms may \textit{not} behave like the thermodynamic or hydrostatic pressure, especially when we consider generalized fluid flows with non-negligible viscosity: the diagonal elements of $\sigma$, which are sometimes called the `mechanical pressure' $p_{\text{mech}}$, will generally not be equivalent to the hydrostatic pressure, differing by terms depending on the substance's viscosity which in turn depends on how particular substances respond to shear stresses.\footnote{See e.g. \citet[\S2.4.3]{white_viscous_1991}, \citet[p. 84]{mihalas_foundations_1984}, or \citet[p. 30]{birkhoff_hydrodynamics_2015}.} 

If so, there are three ways we can `carry on' with the concept of pressure in continuum mechanics. We might either default to Cauchy's claim that the `pressure' -- generalized (normal \textit{and} shear) stresses -- really is \textit{vectorial} and possibly \textit{anisotropic}, or conclude that the hydrostatic pressure -- the isotropic scalar magnitude of balanced forces -- is \textit{undefined}. We can also insist that the normal stresses of $\sigma$ -- $p_{\text{mech}}$ -- `really' is the pressure. In all cases, though, the new pressure loses its physical meaning because it is no longer consilient with the hydrostatic pressure, and, in turn, is no longer consilient with the thermodynamic pressure.  

\y{However, when viscosity becomes significant (e.g., under turbulence or high-speed flow), local equilibrium can fail to hold, equilibrium thermodynamics simply no longer applies, and we might not have reasons to assume $p_{\text{classical}}$ can be defined: steady-state flows may not persist long enough for thermodynamic reasoning to work. In contrast, wherever thermodynamics \emph{does} apply, the four perspectives converge on a single classical pressure, $p_{\text{classical}}$. In this case, consilience holds precisely within the domain of equilibrium thermodynamics, and the breakdown of consilience -- and $p_{\text{classical}}$ -- thus signals the limits of thermodynamics’s applicability.} I suggest that a similar story should be told for the generalization of thermodynamics to the relativistic regime. 

\section{Putting pressure under pressure}

The four perspectives demonstrate remarkable consilience in defining $p_{\text{classical}}$ in the domains where thermodynamics applies: regimes in which systems are (globally or locally) in equilibrium. Extending thermodynamics -- and $p_{\text{classical}}$ -- to render it compatible with the principle of relativity might now seem fairly straightforward. Unfortunately, and perhaps surprisingly, $p_{\text{classical}}$ demonstrates no consilience when we consider relativistic counterparts to the four aforementioned perspectives. Each perspective tells us a different story about how the relativistic pressure ought to behave.

\subsection{Forces over area: relativistic pressure as invariant}

\citet[\S13]{einstein_relativity_1907}, in his quest to find the Lorentz transformation for the pressure, adopted the hydrostatic and classical mechanical perspective by understanding pressure simply as the magnitude of forces over area. From this perspective, he concluded that the relativistic generalization of pressure is that of a Lorentz-invariant quantity: 
\begin{equation}
    p' = p
\end{equation}
Everything is as it were in classical physics; the relativistic pressure retains the isotropy and scalar nature of $p_{\text{classical}}$. 

\begin{figure}[h]
    \centering
    \includegraphics[scale = 0.3]{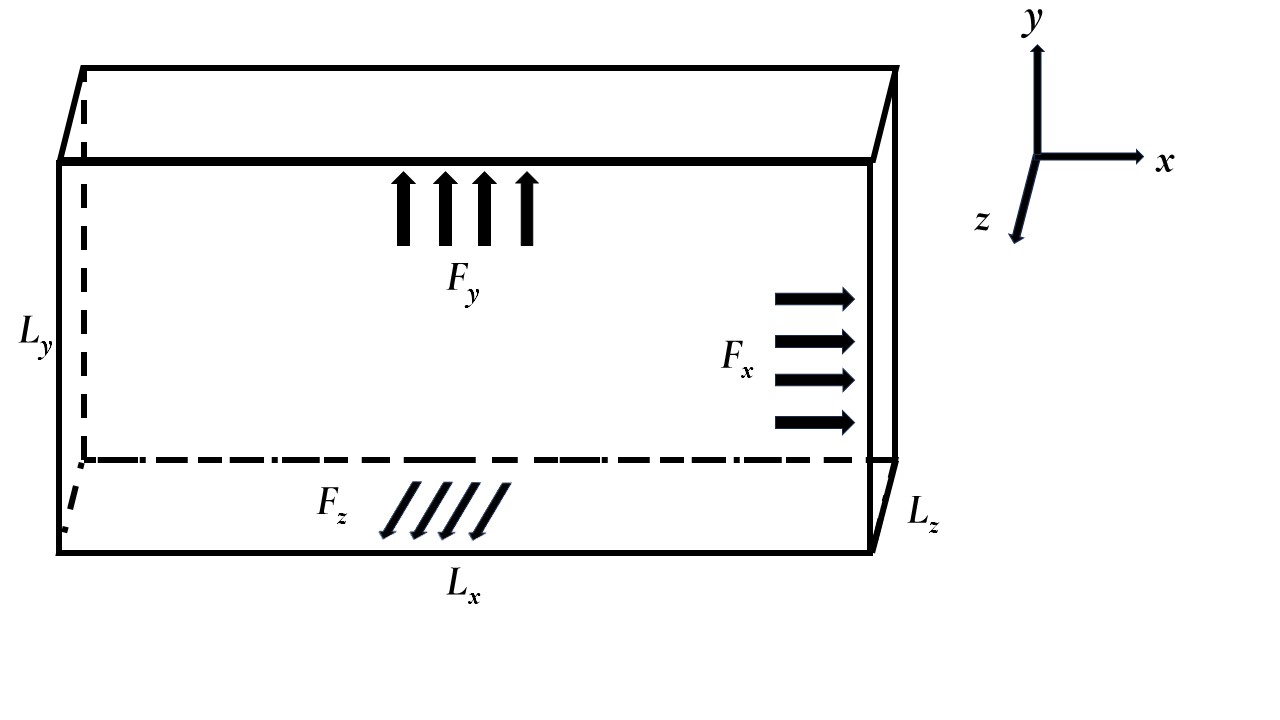}
    \caption{The ingredients for proving the Lorentz invariance of pressure.}
    \label{fig:3}
\end{figure}

Einstein derives the Lorentz-invariance of pressure using the electromotive forces acting on charges. Consider an electric charge bounded by three (electrically charged) one-dimensional elements $L_x$, $L_y$, and $L_z$ each oriented in the respective directions (like a box), in a frame where the charge is at rest, with the electromotive force $\textbf{F} = F_x, F_y, F_z$ exerted on the charge normal to the appropriate 2-dimensional surfaces. That is, $F_x$ acts on the plane bounded by $L_y$ and $L_z$, $F_y$ on $L_xL_z$, and $F_z$ on $L_xL_y$ (see Fig. \ref{fig:3}). 

We know how the electromotive force behaves via the (crucially, Lorentz-invariant) Lorentz force law, which describes the force exerted on a charged particle moving through an electric $ \mathbf{E}$ and magnetic field $\mathbf{B}$:
\begin{equation}
\mathbf{F} = q(\mathbf{E} + \mathbf{v} \times \mathbf{B}) 
\end{equation}
where $q$ is the electric charge of the particle, $\mathbf{E}$ is the electric field, $\mathbf{v}$ is the velocity of the particle, and $\mathbf{B}$ is the magnetic field. The components of this force will transform under a Lorentz boost in the $x-$direction as:\footnote{See e.g. \citet[eq. 21, p. 294]{einstein_relativity_1907}.}
\begin{equation}
     F_x' = F_x, \; \; \; \; F_y' = \frac{1}{\gamma} F_y, \; \; \; \; F_z' = \frac{1}{\gamma} F_z
\end{equation}
We also know how lengths in the rest frame will transform under a Lorentz boost to the moving frame (i.e. length contraction): 
\begin{equation}
    L_x' = \frac{1}{\gamma}L_x,  \; \; \; \; L_y' = L_y,  \; \; \; \; L_z' = L_z
\end{equation}
Finally, we have a definition of pressure in the rest frame, borrowing from the isotropy of pressure for hydrostatic systems from \S3.1: 
\begin{equation}\label{forceoverarea}
p_x = \frac{F}{A} = \frac{F_x}{L_y L_z}, \;  \; \; \; p_y = \frac{F_y}{L_x L_z}, 
 \; \; \; \; p_z = \frac{F_z}{L_x L_y}
\end{equation} 
But now we have all we need to consider the Lorentz transformation of pressure from this perspective. The Lorentz-boosted pressure, $p'_i$ in the $i$\textsuperscript{th} direction is given by:
\begin{equation}\label{pressurefromforce}
p'_x  = \frac{F'_x}{L'_y L'_z} = \frac{F_x}{L_y L_z} = p_x,  \; \; \; \; \;
p'_y  = \frac{F'_y}{L'_x L'_z} = \frac{\gamma F_y}{\gamma L_x L_z} = p_y,  \; \; \; \; \; p'_z  = \frac{F'_z}{L'_x L'_y} = \frac{\gamma F_z}{\gamma L_x L_y} = p_z 
\end{equation}
The length contraction associated with each plane on which the forces act exactly cancels out the Lorentz transformation of the force, and so:
\begin{equation}
    p'_x = p'_y = p'_z = p_x = p_y = p_z
\end{equation}
Thus $p$ is Lorentz-invariant and isotropic. 

However, when we adopt a more contemporary approach in terms of the 4-force, we find Einstein's approach reproduced with an interesting peculiarity. The relativistically appropriate way to understand a generic force is in terms of the 4-force, since the 4-force -- and not the 3-force! -- \x{is a 4-vector that transforms as a tensor (of rank 1) under Lorentz transformations.} The 4-force is generically:
\begin{equation}\textbf{F} = \frac{d\textbf{P}}{d\tau} = (\frac{\gamma dE}{dt}, \frac{\gamma d\textbf{p}}{dt}) \end{equation}
where $\tau$ is the proper time, $\textbf{P}$ is the 4-momentum, $\textbf{F}$ is the 4-force which is the proper time derivative of $\textbf{P}$, $E$ is the energy of the system, and $\textbf{p}$ is the 3-momentum. Suppose we started in the rest frame of some system in a box, in which it is in equilibrium. The system experiences no energy flux over time (``power") in this frame since it is in equilibrium, so:
\begin{equation}F^0 = \frac{\gamma d E}{d t} = 0\end{equation}
Furthermore, the system's velocity in this frame is $v = 0$, $\gamma = 1$, and the 4-force acting on the system is simply:
\begin{equation} \textbf{F} = (0, \frac{d\textbf{p}}{dt})\end{equation}
which we can rewrite in terms of the components of the 3-force since forces are classically equivalent to changes in momenta per Newton's Second Law:
\begin{equation}\textbf{F} = (0, F_x, F_y, F_z) \end{equation}
These are simply the components of the classical force acting in the $x$, $y$, and $z$ directions respectively, representing the forces acting on the box due to the system. We can now ask how this 4-force, represented in the rest frame, transforms under a Lorentz transformation $\Lambda^\lambda_\mu$ into a frame moving in the $x$-direction with velocity $\textbf{u} = (u_x, 0, 0)$. The Lorentz transformation for this simple boost in the $x$-direction is:
\begin{equation}\label{Lorentzmatrix}
\Lambda = \begin{pmatrix}
\gamma & -\gamma u_x & 0 & 0 \\
-\gamma u_x & \gamma & 0 & 0 \\
0 & 0 & 1 & 0 \\
0 & 0 & 0 & 1 \\
\end{pmatrix}
\end{equation}
It follows that:
\begin{equation}F'^\lambda = \Lambda^\lambda_\mu F^\mu \end{equation}
\begin{equation}
    F'^0 = \gamma\frac{dE'}{dt'} = -\gamma u_x F_x,  \; \; \; \; F'^1 = \gamma F_x,  \; \; \; \; F'^2 = F_y,  \; \; \; \; F'^3 = F_z.
\end{equation}
Since in the new frame:
\begin{equation}
    \textbf{F}' = \frac{d\textbf{P}'}{d\tau} = (\frac{\gamma dE'}{dt'}, \gamma F'_x, \gamma F'_y, \gamma F'_z)
\end{equation}
We can compare the terms and conclude that:
\begin{equation}
\begin{aligned}
     F'^0 & = \frac{\gamma dE'}{dt'} = -\gamma u_x F_x \iff \frac{dE'}{dt'} = -u_x F_x \\
    F'^1 & = \gamma F'_x = \gamma F_x \iff F'_x = F_x \\
    F'^2 & = \gamma F'_y = F_y \iff F'_y = \frac{1}{\gamma} F_y \\
    F'^3 & = \gamma F'_z = F_z \iff F'_z = \frac{1}{\gamma} F_z
\end{aligned}
\end{equation}
We get Einstein's results for the 3-force again, which we can use in \eqref{forceoverarea} and \eqref{pressurefromforce} to again conclude the Lorentz-invariance and isotropy of pressure, just like $p_{\text{classical}}$. 

However, the 4-force formalism reveals something elided in Einstein's presentation, the $F'^0$ term, the energy flux or relativistic \textit{power}, which can be interpreted as work being done by the system to its environment, since:
\begin{equation}
    \frac{\gamma dE'}{dt'} = -\gamma F_x u_x \iff dE' = -F_x u_x dt' = -F_x dx' = -dW'
\end{equation}
where $-dW'$ is the work being done by the system in the moving frame on its environment. This transformation suggests that a system in equilibrium in its rest frame, not experiencing any energy flux (it is closed), will appear to experience an velocity-dependent energy flux in a moving frame. It looks like \textit{the system is not in equilibrium after all}; it is losing energy by exerting work on its environment in the moving frame, despite not doing so in the rest frame.

One obvious solution arises when we observe that \textit{the environment} must be interacting with the system in the rest frame, for the system to be in equilibrium. Recall that the pressure must be defined in terms of \textit{balanced forces}. We should therefore also consider how the forces acting on the system due to the environment, e.g. the box, transform. Specifically: in the rest frame, the forces acting on the system due to the box is equal and opposite, on each side of the box, to the forces acting on the box due to the system which we've just calculated. Since the box, in the rest frame, is also not exchanging energy with the system, we can associate with it a 4-force that is exactly opposite to that of the system:
\begin{equation}\textbf{F} = (0, -F_x, -F_y, -F_z) \end{equation}
Again, the force over surface area will be Lorentz-invariant, but the energy flux of the box will be \textit{equal but opposite} to that of the system:
\begin{equation}
    F'^0_{\text{box}} = F_x u_x \iff  dE' = F_x u_x dt' = F_x dx' = dW'
\end{equation}
Thus, to a stationary observer, the system -- \textit{sans box} -- in the moving frame will appear to be losing energy and doing work to its environment. But that's an \textit{incomplete description} of the system in equilibrium; what made it possible to ascribe an equilibrium pressure -- and generally, an equilibrium state -- to the system was precisely the existence of a box (or more generally, an adjacent system) acting with equal and opposite forces to the system. Crucially, in the moving frame, the box will also appear to be doing work to the system with equal magnitude but opposite direction, such that the overall system -- the box \textit{and} system inside it -- remains in equilibrium. 

This suggests a need for caution with our unit of analysis in relativistic thermodynamics. In classical thermodynamics we can often ignore the box when discussing forces; the box is simply acting on the system with forces of equal and opposite magnitude, so specifying the forces acting on the box by the system suffices. But upon a Lorentz boost, the box -- and the effects of the Lorentz boost on the box -- \textit{must also be included in the description to properly characterize the system in thermodynamic equilibrium}. As \citet{penney_note_1966} emphasizes: ``it is always assumed to be no unbalanced forces acting
during a quasi-static process, aside from infinitesimals of high order. \textit{One cannot think of an external pressure as being applied without an internal pressure to balance it almost completely}." 

\subsection{The fundamental equation: relativistic pressure increases for the moving observer}

Einstein and Planck stop here, concluding that the relativistic pressure is Lorentz-invariant. Given $\S3$, though, it's clear that the approach in terms of the magnitude of forces over area is merely one out of various perspectives we can take in order to understand $p_{\text{classical}}$, and hence how it extends into relativity. For instance, when we consider how thermodynamic quantities such as $p_{\text{classical}}$, \textit{defined in terms of the fundamental equation}, transforms under Lorentz transformations, we find a breakdown of consilience with the perspective from \S4.1. To get there, we must confront two problems at the foundations of relativistic thermodynamics. 

\subsubsection{What is the relativistic fundamental relation?}

The first problem concerns the subtle question of \textit{the fundamental relation}. That is, what should the form of the fundamental relation be, for systems in a moving frame? Prominently, we've seen that the internal energy of a system \textit{alone} \textit{is not conserved} when Lorentz-boosted, contrary to what we might think classically. This means that we cannot necessarily just borrow the classical fundamental relation
\begin{equation}
dU = TdS - pdV + \mu dN
\end{equation}
which describes all the ways that $U$ can change due to thermodynamic processes. However, this does not appear to exhaust the possible ways in which internal energy can change for a relatively moving system. \citet[p. 199]{liu_einstein_1992} discusses von Laue's proposal in the context, that we now need
\begin{quote}
    an amount of work, namely, that which is necessary to enable the moving body to release heat with constant velocity. Because of that its momentum is known to be changed. In order to sustain the velocity a force is thus necessary, which also produces work.
\end{quote}
On this view, even if the system is at rest -- and in equilibrium -- with respect to another body while exchanging heat (i.e. energy) with that body in the rest frame, the same heat exchange, in the moving frame, results in a loss of mass due to $E = mc^2$, which leads to a change in linear momentum equivalent to $-\textbf{u}.d\textbf{G}'$.\footnote{See \citet[p. 188]{liu_einstein_1992}.} This leads to \textit{deceleration} -- disequilibration -- and apparent paradox: the two frames no longer describe the same physical phenomenon! Hence von Laue's suggestion that we need to include an additional work term, acting on the system, equal but opposite to $\textbf{u}.d\textbf{G}'$, in order to counteract this deceleration and ensure that the system is still in thermodynamic equilibrium with the other body. This is the `translational work' term, exactly equal (but opposite) to the deceleration.

In principle, we can add other pairs of intensive parameters $X$ and extensive parameters $Y$ to the fundamental relation, such that: 
\begin{equation}dU = TdS - pdV + \mu dN + X dY ... \end{equation}
defining $X$ in terms of the partial derivative $\frac{\partial U}{\partial Y}|_{S, V, N...}$, that is, so long as $Y$ can be varied while \textit{holding all other extensive parameters fixed}. Hence, \citet{planck_dynamics_1908} proposed to include in the fundamental relation the `translational work' term, comprising of the intensive-extensive parameter pair $\textbf{u}.d\textbf{G}$, where $\textbf{u}$ is the system's velocity, and $\textbf{G}$ is the momentum of the system, such that the fundamental relation in the moving frame is now:
\begin{equation}\label{planckianFR}
dU' = T'dS' - p'dV' + \mu' dN' + \textbf{u}.d\textbf{G}' \end{equation}
which exactly compensates for the deceleration. Including this `translational work', $dU$' is conserved, and we can use this fundamental relation for thermodynamic reasoning as before. Call this the \textit{Planckian} perspective. On this view, additional work needs to be put into a system in a frame moving with velocity $\textbf{u}$, proportional to $\textbf{u}.d\textbf{G}'$, to even enable the possibility of heat transfer between systems in equilibrium. Otherwise the system shifts out of equilibrium due to deceleration and heat transfer at constant velocity is impossible. The translational work is a \textit{precondition} for heat transfer at constant velocity $\textbf{u}$.

Curiously, while \citet{einstein_relativity_1907} initially followed a similar line of reasoning, he later changed his mind and argued against von Laue's reasoning in letters from 1952--1953:\footnote{See \citet{liu_einstein_1992, liu_is_1994, chua_t_2023}.}
\begin{quote}
    When a heat exchange takes place between a reservoir and a `machine' - both of them are at rest with each other and acceleration-free, it does not require work in this process. This holds independently whether both of them are at rest with respect to the employed coordinate system or in a uniform motion relative to it. (\cite[fn. 31]{liu_einstein_1992})
\end{quote}
Einstein seems concerned about the need to introduce an additional thermodynamic process -- manipulating the `translational work' -- in order to enable heat transfer between two systems \textit{known to be in equilibrium} (at rest with each other) in their rest frame. 

I am tempted to agree with Einstein here. The translational work term, as it relates to heat transfer, is conceptually unclear to me, because it's generally unclear where this `work' comes from. \citet{liu_einstein_1992} mentions that the heat has to ``contribute an extra amount of its content to doing work'', but what does this `contributing' process amount to? ``Work" is supposed to be a prerequisite for the possibility of heat transfer, an extra work we need to put in by hand \textit{in order to} facilitate heat exchange between two systems. However, from the two systems' perspectives, they are in relative equilibrium and there is only ever \textit{one thermodynamic process}: heat transfer (suppose at fixed $V$ and $N$). In the rest frame, one system is simply losing energy via heat, and the other is gaining energy via heat. In the moving frame, the same system is now losing energy \textit{plus momentum} via $\textbf{u}.d\textbf{G}'$, while the same body is now gaining energy \textit{plus momentum} via $\textbf{u}.d\textbf{G}'$. Where, then, is the new thermodynamic process in terms of `translational work' supposed to come in? \citet{planck_dynamics_1908} simply asserts the existence of this translational work term, while von Laue seems to understand this work as a prerequisite \textit{for} the transfer of heat ``necessary to enable the moving body to release heat with constant velocity" -- but doesn't explain where it comes from.  

\begin{figure}[h]
    \centering
    \includegraphics[scale = 0.3]{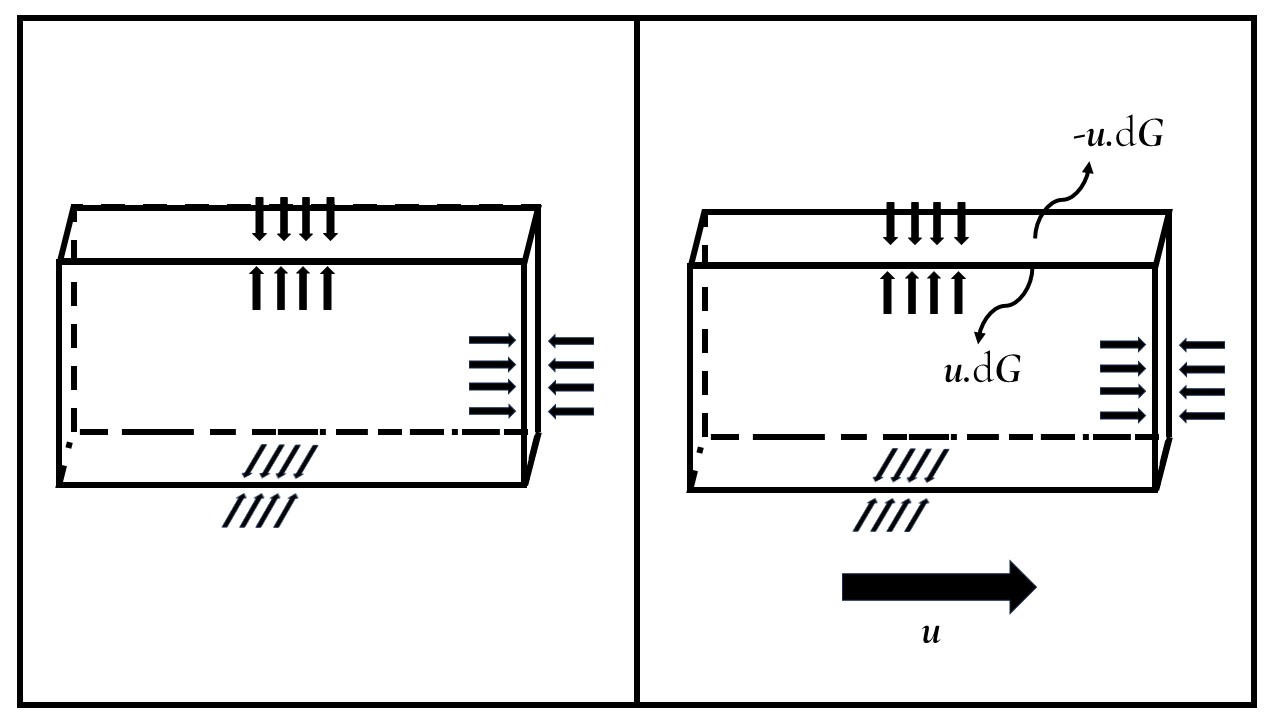}
    \caption{\textbf{Left}: a system having equilibrated with its box, without energy flux. \textbf{Right}: the same box from the perspective of a Lorentz-boosted frame, with constant velocity $\textbf{u}$.}
    \label{fig:4}
\end{figure}

An alternative understanding of the translational work seems much clearer. Call this the \textit{Einsteinian} perspective: we `include the box' (where the `box' can be a generic environment), and furthermore we generalize our understanding of thermodynamic equilibrium. From \S4.1, we saw that a system at rest has to have a balance of forces with its environment to be in mutual equilibrium. In the moving frame, the forces acting on the box must \textit{also} be balanced with its environment for them to be in mutual equilibrium (and more importantly, to be considered to be in \textit{constant motion} without deceleration/acceleration). For this balance to obtain, there must \textit{not} only be a balance of the usual incoming forces, but also a balance of \textit{an additional quantity} due to the energy flux -- and hence work and hence an additional force term -- coming from the box. Recall from \S4.1 that \textit{any} system, experiencing no energy flux and some forces in its rest frame, will always appear in the Lorentz-boosted frame to experience some energy flux. (See Fig. \ref{fig:4}.) 
Crucially, this energy flux is \textit{exactly equivalent to} $-u_x.F_x$, where $F_x$ is the force along the direction of motion of the moving frame. This is, in turn, equivalent to the translational work $-\textbf{u}.d\textbf{G}'$ where $\textbf{u} = (u_x, 0, 0)$ since $\textbf{G}$ is just 3-momentum $\textbf{p}$ by another name. Furthermore, once we `include the box', $-\textbf{u}.d\textbf{G}'$ is canceled out. This suggests a generalized understanding of thermodynamic equilibrium: \textit{the balance of forces must generalize to include a balance of energy flux -- or `translational work' -- between the system and the box}. This new understanding of thermodynamic equilibrium also vindicates Einstein's suggestion that \textit{no extra work needs to be done to facilitate ordinary thermodynamic processes}. For the case of pressure exchanges, the forces are balanced in the rest frame; in the moving frame, the forces \textit{plus energy flux} are balanced. 

This reveals the source of the translational work term: it's neither a new kind of thermodynamic process to be `put in by hand', nor a new means of manipulating a system thermodynamically (unlike the other intensive-extensive parameter pairs). Rather, it arises because we have `ignored the box'. The apparent paradox of deceleration due to heat transfer -- and the supposed need to do `translational work' -- comes from failing to recognize that the co-moving environment already needs to supply an equal but opposite amount of work to the system as it receives heat from the system, if both are to remain in equilibrium.  More generally, from this Einsteinian perspective, there's no need to add in an extra `translational work' term when we `include the box', since no extra work ever needs to be done. For systems in equilibrium, the usual thermodynamic processes \textit{already include this exchange of energy-plus-momentum} in what it means to exchange ordinary thermodynamic quantities e.g.  heat or pressure. 

Furthermore, this generalized understanding of equilibrium and equilibrium thermodynamic processes comes with clarity about the fundamental relation. Classically, what we typically take to be conserved for a system in equilibrium is the \textit{system}'s internal energy $U$. However, if we `ignore the box', systems can appear to experience an energy flux as long as it's moving with constant velocity relative to a stationary observer, even when no other thermodynamic processes are occurring, such that:
\begin{equation}
    dU_{\text{system}}' = - \textbf{u}.d\textbf{G}'
\end{equation}
If we insist on focusing on the fundamental relation for the system in terms of $U'_{\text{system}}$ as a conserved quantity, we must have a correction via $\textbf{u}.d\textbf{G}'$ such that for a closed system
\begin{equation}
    dU_{\text{system}}' + \textbf{u}.d\textbf{G}' = 0
\end{equation}
This suggests that $\textbf{G}'$ is an extensive parameter, leading to the Planckian perspective.

The Einsteinian perspective instead tells us to ``include the box" when we are considering Lorentz-boosting a thermodynamic system, and to reconsider what is balanced in equilibrium for moving systems. Since the energy flux $-\textbf{u}.d\textbf{G}'$ leaving the system is balanced by the energy flux $\textbf{u}.d\textbf{G}'$ from the box in the moving frame, then when no other thermodynamic processes are occurring, 
\begin{equation}
    dU_{\text{system}}' - \textbf{u}.d\textbf{G}' + dU_{\text{box}}' + \textbf{u}.d\textbf{G}' = dU_{\text{system}}' + dU_{\text{box}}' = 0.
\end{equation}
That is, the form of the fundamental relation \textit{does not change} in the moving frame -- we do not need to add an extra `translational work' term to $dU'$. It is true that the usual thermodynamic processes, i.e. exchanges of heat \textit{and} work, which only involve exchanges of energy in the rest frame can involve exchanges of energy \textit{and} momentum in general for moving frames, even from the Einsteinian perspective. From the Einsteinian perspective, however, such exchanges are already included into what it means for a system to undergo the usual thermodynamic processes via changes in e.g. $S, V, N$, and there is no need to include an additional correction term or a new extensive-intensive parameter pair. So the fundamental relation, even in the moving frame, is still understood only in terms of the usual thermodynamic variables in the moving frame:
\begin{equation}\label{einsteinianFR}
    dU' = T'dS' - p'dV' + \mu' dN'
\end{equation}
I much prefer the Einsteinian perspective over the Planckian one.\footnote{See also \citet[p. 29--30]{landsberg_a_1967} and their `confined'/`inclusive' distinction.} The Planckian perspective is useful for seeing how exactly the system per se behaves. However, the view can obscure important features of the system, namely, interaction and equilibration with its environment.

\subsubsection{What is the Lorentz transformation for energy?}

We are now finally in a position to ask how $p_{\text{classical}}$ generalizes to special relativity from the perspective of both relativistic fundamental relations: unsurprisingly, the two perspectives disagree about how the relativistic pressure transforms. Furthermore, both perspectives tell us something \textit{quite different} from the mechanical perspective from \S4.1. 

From the Planckian perspective, the fundamental relation in the moving frame is the sum of the internal energy and translational work, i.e. \eqref{planckianFR}. An alternative is to hold fast, as later Einstein did, to the classical thermodynamic relation, treating the `translational work' term to be entirely superfluous to what's thermodynamically relevant to the system: it's already included in what it means for pressure or heat to be exchanged with another system in equilibrium. For a relatively moving system, its fundamental relation -- capturing all that's thermodynamically relevant to the system -- is simply \eqref{einsteinianFR}. In both cases, to define the pressure using the thermodynamic perspective, we simply hold all other extensive variables fixed and consider:
\begin{equation}
\frac{\partial U'}{\partial V'} = -p' \end{equation} 
Now consider how $U$ and $V$ transform. We already know that the volume contracts, due to length contraction, such that, for a system moving at constant velocity $u$ in the $x$-direction, i.e. with a constant magnitude $u_x$:
\begin{equation}\label{volumetrans} dV' = \frac{1}{\gamma}dV
\end{equation}
so what remains is how $U$ transforms. However, a second major problem shows up, surrounding the Lorentz transformation of \textit{energy}, which \textit{also} rests on the perspective one takes on the unit of analysis for relativistic thermodynamic systems. 

The pioneers of relativistic thermodynamics and those that followed soon after, such as \citet{planck_dynamics_1908, tolman_relativity_1934, pauli_theory_1958}, as well as some in the 1960s debate such as \citet{kibble_relativistic_1966}, argued that the total energy $E$ -- and hence the internal energy $U$ -- should transform as:
\begin{equation}\label{planckianenergy}
    U' = \gamma (U + u_x^2 pV) 
\end{equation}
where $p$ and $V$ are the pressure and volume of the system -- crucially, without the box -- \textit{in its rest frame}. This transformation follows directly from the Lorentz transformation of $T^{\mu\nu}$ for a perfect fluid (which I return to in \S4.4). Note, though, that this essentially assumes a Planckian perspective: recall that $\sigma_{ii}$ (and its stress-energy tensor analogue, to be discussed later) represents only the \textit{internal} normal stresses. We are essentially only considering the forces exerted by the system, not the corresponding counterbalancing forces -- normal stresses with opposite direction and equal magnitude -- associated with the box. 

An alternative for the Lorentz transformation of energy comes from the Einsteinian perspective, defended by e.g. \citet{ott_lorentz-transformation_1963, landsberg_a_1967, arzelies_transformation_1965, sutcliffe_lorentz_1965, gamba_relativistic_1965}. We've already seen reasons for considering this: the system without its box (again, more generally, its environment) cannot really be understood to be in equilibrium. However, just as this canceled out the effects of the translational work and kept the system and its environment in equilibrium, it also cancels out the $-u_x^2 pV$ term because the normal stresses exerted by the box acting on the system act with equal magnitude but opposite direction to the normal stresses exerted by the system on the box. Consequently:
\begin{equation}\label{einsteinianenergy}
    U' = \gamma U
\end{equation}
Now, if we adopted the Einsteinian transformation in considering the transformation of pressure, we get:
\begin{equation}\label{pressuretrans}
    p' = -\frac{\partial U'}{\partial V'} = -\frac{\gamma\partial U}{1/\gamma \; \partial V} = \gamma^2 p
\end{equation}
If we instead adopted the Planckian transformation for energy, then:
\begin{equation}U' = \gamma(U + u_x^2pV)\end{equation}
However, on this understanding of energy, combined with our earlier procedure, we find the same relation, together with extra velocity-dependent terms:
\begin{equation}p' = -\frac{\partial U'}{\partial V'} = -\frac{\gamma\partial (U + u_x^2pV)}{1/\gamma \; \partial V} = -\gamma^2 \bigg( \frac{\partial U}{\partial V} + \frac{\partial u_x^2pV}{\partial V} \bigg)= \gamma^2 (p - u_x^2 p - u_x^2V\frac{\partial p}{\partial V})\end{equation}
Yet again, the moving pressure is \textit{generally distinct} from the rest pressure, with the last term depending on the specific equation of state we employ (something not fixed by the fundamental relation per se, as mentioned in \S3.3). This is dramatically different from the earlier understanding of relativistic pressure as a Lorentz invariant concept. While the pressure remains isotropic, it is no longer \textit{Lorentz-invariant}; it transforms as $p' = \gamma^2 p$ or $p' = \gamma^2 (p - u_x^2 p -...)$ from the perspective of the fundamental relation depending on whether we adopt the Einsteinian or Planckian perspective, rather than $p' = p$ from considering the 4-force. 

Some, like \citet{sutcliffe_lorentz_1965}, \citet{farias_what_2017}, \citet{brechet_relativistic_2022}, have already made this observation, distinguishing this new pressure concept (they prefer the Einsteinian $p' = \gamma^2 p$) as the \textit{thermodynamic pressure}, distinct from the \textit{hydrostatic pressure} concept they associate with $p' = p$. For instance, \citet[p. 686]{sutcliffe_lorentz_1965} points out that ``the two pressures are the same only in the local rest frame. This distinction is necessary in order for relativistic thermodynamics to be consistent with relativistic mechanics." Likewise, \citet[p. 7]{farias_what_2017} observes that
\begin{quote}
    ...this discrepancy [between Lorentz-invariant $p$ and the $p$ we obtained from the fundamental relation] is related to the different definitions of pressure. In our approach, that identifies intensive quantities as the ones that are equal between two subsystems in thermodynamic equilibrium... In the case of pressure, it is only equal to force divided by area in the rest frame of the system $A$. 
\end{quote}
\citet[p. 35]{brechet_relativistic_2022} similarly observes that ``the thermodynamic pressure ... has to be distinguished from the mechanical pressure that is invariant under a relativistic frame transformation." 

However, the bulk of the literature on relativistic thermodynamics has focused on the temperature. Indeed, debates about the energy transformation entirely centered on the upshot for temperature, with most people assuming that `the' pressure is simply the hydrostatic pressure. One outlier is \citet{balescu_relativistic_1968}, who explicitly entertains the possibility of the relativistic pressure transforming as $p' = \gamma^2 p$, attributing it to \citet{sutcliffe_lorentz_1965}. However, he rejects it quickly without argument:
\begin{quote}
    [Sutcliffe] realized actually that the pressure defined by the usual [\eqref{pressuretrans}] from [$E$] cannot be identified with the scalar hydrodynamical pressure. As a result he proposed to make a distinction between the mechanical pressure and the thermodynamic pressure... Such a distinction has no physical basis whatever. (\cite[p. 328]{balescu_relativistic_1968})
\end{quote}
But this seems wrong. We've already seen that the mechanical pressure is conceptually distinct -- and diverges -- from the thermodynamic pressure in general continuum mechanical contexts, despite their remarkable consilience in the hydrostatic context. They already `fall apart' in other settings. 

What makes the present situation initially worrisome is the lack of reason yet to question the application of thermodynamics here: in the system and environment's own rest frame, the frame in which they are both at rest with respect to each other, thermodynamics is business as usual. However, when we want to continue our thermodynamic analysis in other frames in accordance with the principle of relativity, we again find divergence between the new relativistic `pressure' and $p_{\text{classical}}$. Perhaps, as in the general continuum mechanical context, this suggests that we have crossed the limits of applicability of thermodynamics when we try to apply thermodynamics to systems in relatively moving frames. In these cases, there might be no natural answer about the thermodynamic state of that system because $p_{\text{classical}}$ breaks down and loses consilience.

\subsection{Equations of state: pressure meets relativistic temperature}

Recall that the equations of state are what fix the particular thermodynamic behavior of a system, such as phase transitions. One thing we might consider, then, is to hold fixed the specific form of the equations of state for a system, both in its rest frame and moving frame. For instance, for a box of ideal gas, we might demand that in the rest frame, the Ideal Gas Law holds:
\begin{equation}
    pV = Nk_BT
\end{equation}
and also in the moving frame:
\begin{equation}
p'V' = Nk_BT' 
\end{equation}
where we assume that the particle number $N$ is a Lorentz-invariant quantity. Then, the question is whether we can ascertain the transformation of $p'$ in terms of $p$ given the above, something \citet{sutcliffe_lorentz_1965} proposed as a desideratum. 

However, we immediately face the arbitrariness and `falling apart' of the relativistic temperature, leading to the same arbitrariness and `falling apart' of the relativistic pressure. As mentioned in \S2, there are three proposals for the Lorentz transformation of temperature, the Planck-Einstein temperature:
\begin{equation}
T' = \frac{1}{\gamma} T 
\end{equation}
The Einstein-Ott temperature:
\begin{equation} 
T' = \gamma T 
\end{equation}
and the Landsberg temperature: 
\begin{equation} 
T' = T 
\end{equation}
Starting with 
\begin{equation}
p'V' = Nk_BT' 
\end{equation}
and again invoking the Lorentz transformation for volume i.e. \eqref{volumetrans}, three corresponding proposals emerge.  Given the Planck-Einstein temperature,
\begin{equation} 
p' = p.
\end{equation}
Given the Einstein-Ott temperature,
\begin{equation}
p' = \gamma^2 p.     
\end{equation}
And given the Landsberg temperature, we get \textit{yet} another proposal for the relativistic pressure:
\begin{equation}
    p' = \gamma p.   
\end{equation}
The `Landsberg pressure' defined this way still appears to increase for a system in the moving frame. However, it increases less -- by a factor of $\gamma$ -- than the `Einstein-Ott pressure'. 

Furthermore, we've only considered the simplest case of the equations of state for an ideal gas. Even a slightly more realistic equation of state, such as Van Der Waal's equation of state, brings in other complications because they don't have a simple functional form. For instance, we could demand that the Van Der Waal's equation of state also holds in the moving frame:
\begin{equation}
    p' = \frac{RT'}{V'_m-b'}-\frac{a}{V_m'^2}.
\end{equation}
Since $b$ is the excluded volume due to the gas molecules' finite size, we expect it to also transform like a volume $V' = \frac{1}{\gamma}V$ so
\begin{equation}
    p' = \frac{\gamma RT'}{V_m-b}-\frac{\gamma^2 a}{V_m^2}.
\end{equation}
If $T' = \gamma T$, then, again,
\begin{equation}
    p' = \gamma^2 p.
\end{equation}
However, if we adopted either $T' = \frac{1}{\gamma}T$ or $T' = T$, it's easy to see that no simple relationship between $p$ and $p'$ may be found. 

From this perspective, we have to return to the debate over relativistic temperature. Yet, as \citet{chua_t_2023} discusses, it might \textit{also} be difficult to say anything conclusive there. This renders the status of relativistic pressure, from this perspective, arbitrary. Insofar as the equations of state are what connect the abstract thermodynamic framework to empirical observations, a failure to render the relativistic pressure compatible with them entails a loss of empirical meaning for the relativistic pressure.

\subsection{Continuum mechanics: anisotropic pressure}

Finally, we come to the relativistic counterpart of continuum mechanics. In many ways, this is intricately connected to the framework of \textit{general relativity},\footnote{See also \citet{doi:10.1086/737745} for further discussion of how temperature as a concept fragments in general relativity, in addition to the worries raised by \citet{chua_t_2023} for special relativity.} where the 3-dimensional stress tensor is replaced by a 4-dimensional stress-energy tensor $T^{\mu\nu}$:
\[
\textbf{T} = 
\begin{bmatrix}
T^{00} & T^{01} & T^{02} & T^{03} \\
T^{10} & T^{11} & T^{12} & T^{13} \\
T^{20} & T^{21} & T^{22} & T^{23} \\
T^{30} & T^{31} & T^{32} & T^{33} 
\end{bmatrix}
\]
In general relativity, $T^{\mu\nu}$ is the source term in Einstein's field equations, which constrains how the metric tensor behaves.  

$T^{\mu\nu}$ now includes four types of terms: stresses, energy density, and energy/momentum density flux. Specifically, $T^{00}$ represents the energy density $\rho$. The components $T^{0i}$ and $T^{i0}$ (with $i = 1, 2, 3$) describe the flux of energy or momentum density in the $i$-th spatial direction (with 1, 2, 3 corresponding to the $x$, $y$, and $z$ directions, respectively; 0 represents time). The remaining components correspond to the Cauchy stress tensor $\sigma$ (up to sign difference). The diagonal components $T^{ii}$ represent normal stresses in the $i$-th directions, while the off-diagonal components $T^{ij}$ (for $i \neq j$) represent shear stresses.

We can consider hydrostatic systems using a stress-energy tensor for \textit{perfect fluids}. \x{For instance, a common example of a perfect fluid is a photon gas, with equation of state $p = \rho/3$.} In terms of a frame where the \x{photon gas} is at rest, $T^{\mu\nu}$ is:
\begin{equation}
\begin{bmatrix}
\rho & 0 & 0 & 0 \\
0 & p & 0 & 0 \\
0 & 0 & p & 0 \\
0 & 0 & 0 & p \\
\end{bmatrix}
\end{equation}
In this frame, the system is in equilibrium, with no energy flux. Again, $p$ represents the magnitude of balanced forces acting by/on this system, which happens to be isotropic for reasons already discussed in \S3.1. We can now ask how this system appears under Lorentz transformations to a moving frame. We compute
\begin{equation} T'^{\mu\nu} = \Lambda^\mu_\alpha \Lambda^\nu_\beta T^{\alpha\beta}\end{equation}
using $\Lambda$ from \eqref{Lorentzmatrix}. $T'^{\mu\nu}$ are the components of the stress-energy tensor of the Lorentz-boosted system. We get:
\begin{equation}
\textbf{T}' = 
\begin{bmatrix}
\gamma^2 \rho + u_x^2\gamma^2p & -u_x\gamma^2\rho - u_x\gamma^2 p & 0 & 0 \\
-u_x\gamma^2\rho - u_x\gamma^2 p & u_x^2\gamma^2\rho + \gamma^2 p & 0 & 0 \\
0 & 0 & p & 0 \\
0 & 0 & 0 & p \\
\end{bmatrix}
\end{equation}
This appears to vindicate the Planckian Lorentz transformation of energy, i.e. \eqref{planckianenergy}. Consider
\begin{equation}\label{rhoprime}
\rho' = T'^{00} = \gamma^2 \rho + u_x^2 \gamma^2p
\end{equation}
where $\rho$ is the energy density of the system at rest, and $\rho'$ the energy density of the system in the moving frame. The total energy is the density multiplied by volume, so
\begin{equation}
    U' = \rho' V',\; \; U = \rho V
\end{equation}
We multiply \eqref{rhoprime} by $V' = \frac{1}{\gamma}V$ such that:
\begin{equation}
    U' = \rho'V' = \gamma(U + u_x^2pV)
\end{equation}
which is just \eqref{planckianenergy}. However, we also find that the transformed pressure is \textit{not} isotropic, unlike what Planck and others holding the Planckian perspective had assumed. We can see this by considering the diagonal components of $T'$: 
\begin{equation}
\begin{aligned}
T'^{11} & = u_x^2\gamma^2\rho + \gamma^2 p \\
& \neq T'^{22} = T'^{33} = p    
\end{aligned}
\end{equation}
The magnitude of the normal stress in the $x$-direction is clearly \textit{no longer equivalent} to the magnitude of the normal stresses in the $y$ and $z$ directions. Even if such normal stresses are balanced with equal opposite stresses, \textit{there can no longer be a single scalar quantity} which can capture the magnitude of such balanced forces in the moving frame since the magnitudes of these normal stresses, while isotropic in the rest frame, are \textit{no longer isotropic} in the moving frame, contrary to earlier perspectives.\footnote{The extra $\rho$-dependent term resembles the `moving pressure' or `dynamic pressure' in the Bernoulli equation for moving fluids.}

Furthermore, the same issue of energy flux returns -- in the moving frame, the system appears to be losing energy in the $x$-direction since:
\begin{equation}T'^{01} = T'^{10} = -u_x\gamma^2 (\rho +  p) \end{equation}
and the momentum density flux in the moving frame can be obtained by projecting $\textbf{T}$ onto $\textbf{u}$:
\begin{equation}
    \textbf{T}'\textbf{u}\iff T'^{01}u_x = -u_x^2\gamma^2 (\rho + p)
\end{equation}
which emphasizes that from the Planckian perspective, it is hard to see why the system in the moving frame should even appear to be in thermodynamic equilibrium at all. 

Alternatively, if we again adopted the Einsteinian perspective and `included the box', then we must consider the net normal stresses of the system plus box to be \textit{zero} due to the balanced stresses. Importantly, $T^{\mu\nu}$ for such a system-plus-box is \textit{not} what is typically called $T^{\mu\nu}$ of a perfect fluid, but rather $T^{\mu\nu}$ for \textit{dust}. For the system-plus-box at rest:
\begin{equation}
\textbf{T} = 
\begin{bmatrix}
\rho & 0 & 0 & 0 \\
0 & 0 & 0 & 0 \\
0 & 0 & 0 & 0 \\
0 & 0 & 0 & 0 \\
\end{bmatrix}
\end{equation}
What needs to be Lorentz-boosted is the joint system-plus-box. When we consider the Lorentz transformation of \textit{this} stress-energy tensor, we find that this perspective, unsurprisingly, vindicates the Einsteinian Lorentz transformation for energy, \eqref{einsteinianenergy}. Since:
\begin{equation}\rho' = T'^{00} = \gamma^2\rho\end{equation}
we find that
\begin{equation}U' = \rho'V' = T'^{00}V' = \gamma \rho V = \gamma U \end{equation}
and more generally:
\begin{equation}
\textbf{T}' = 
\begin{bmatrix}
\gamma^2 \rho & -u_x\gamma^2\rho & 0 & 0 \\
-u_x\gamma^2\rho &u_x^2\gamma^2\rho & 0 & 0 \\
0 & 0 & 0 & 0 \\
0 & 0 & 0 & 0 \\
\end{bmatrix}
\end{equation}
This perspective clarifies that the stresses might not be balanced in the Einsteinian picture if we don't generalize our understanding of equilibrium. If we followed convention and \textit{defined} the pressure in terms of the components of $T^{11}$, $T^{22}$, and $T^{33}$, the system appears to experience an additional normal stress in the $x$-direction, in addition to \textit{more anisotropy of pressure}. Even if we assumed that all the normal stresses are balanced in the rest frame, such that the net force is zero, the normal stress in the $x$-direction is \textit{not} balanced. Instead, it is:
\begin{equation}
T'^{11} = u_x^2\gamma^2\rho \neq 0
\end{equation}
while $T'^{22} = T'^{33} = 0$ implying a balance of normal stresses in those dimensions. However, 
\begin{equation}T'^{01} = T'^{10} = -u_x\gamma^2\rho\end{equation}
and so the energy flux in the co-moving frame, projecting onto the co-moving observer, is $-u_x^2\gamma^2\rho$, implying an energy flux (which can also be interpreted as a momentum flux, a force!) exactly equivalent in magnitude -- and opposite in direction -- to the increase in normal stress in the $x$-direction. If we generalize our understanding of equilibrium such that energy-momentum flux are balanced ((i.e. the `translational work' vanishes), in addition to balanced normal stresses, then the system \textit{remains in equilibrium} even in the moving frame. 

From the continuum mechanical perspective, though, both the Planckian and Einsteinian perspectives tell us something interesting about how pressure extends into special relativity. The Planckian perspective tells us that the magnitudes of the normal stresses cease to be isotropic in the moving frame. This is true even for the Einsteinian picture, since the balance of forces only demand that the normal stresses cancel out in all directions, but allows that the \textit{magnitude} of the forces being balanced need not be equal in all directions (and the Planckian perspective tells us exactly how the magnitudes change differently in different directions). Of course, starting from the Einsteinian perspective, we see that the normal stresses starting out balanced also \textit{remain balanced} on a generalized understanding of equilibrium and the balance of forces, even in a moving frame. 

As in general continuum mechanics, on the one hand, if we defined the pressure in terms of the isotropic scalar hydrostatic pressure, then it is undefined when a system is Lorentz-boosted: since the normal stresses are anisotropic, there is no single scalar magnitude which characterizes the balanced normal stresses of the system. On the other hand, if we instead defined the pressure as an anisotropic vectorial quantity, its Lorentz transformations still fail to be consilient with the transformations found via other perspectives.

\subsection{Relativistic pressure as rest pressure?}

The four perspectives provide four incompatible ways for generalizing $p$ to relativity: as a Lorentz-invariant isotropic quantity, as a Lorentz-\textit{non}invariant isotropic quantity, arbitrarily, or as a Lorentz-\textit{non}invariant \textit{an}isotropic quantity. The consilience of the classical pressure breaks down in relativity.  

There is one last way to avoid this situation, akin to \citet{landsberg_a_1967}'s suggestion that we define the Lorentz transformation of temperature to be the temperature \textit{measured in the rest frame} of a system by a co-moving thermometer, even if the system is in relative motion.\footnote{See also \citet[\S4.2]{chua_t_2023}.} Likewise, we can \textit{stipulate} that the relativistic generalization of $p_{\text{classical}}$ is the pressure of a system measured in its rest frame, perhaps by a co-moving pressure gauge, so that the Lorentz transformation of pressure is
\begin{equation}
    p' = p
\end{equation}
where $p$ is defined only in the rest frame of the system. This, however, seems to be a trivial transformation. It amounts to saying that pressure is well-defined where it is well-defined, while ignoring the fact that pressure is \textit{not} uniquely defined in other frames. If so, $p_{\text{classical}}$ does \textit{not} clearly generalize to the relativistic context: for most relativistically allowed frames, that is, frames related by Lorentz transformations, where the system is not at rest, there is no unique thermodynamic description of that system. 

Although unhelpful for generalizing $p_{\text{classical}}$ into the relativistic domain, this option \textit{does} explicitly highlight a limit for the emergence of a thermodynamic description for a system in relativity, namely, when a system is described in the appropriate rest frame. When the system (-plus-environment) is described in a frame such that $\textbf{u} \to 0$, all four perspectives approach consilience again with respect to $p_{\text{classical}}$.

\subsection{Taking stock}

To take stock of the dialectic so far:
\begin{itemize}
    \item Starting with an analysis in terms of normal forces over area, we get the standard result, per Einstein/Planck, that the relativistic pressure is isotropic and Lorentz-invariant: $p' = p$. However, it raises a question about what the appropriate unit of analysis for a system is supposed to be. 
    \item Starting instead with the fundamental relation, we get the result that the relativistic pressure is isotropic but Lorentz \textit{non}-invariant, with options depending on the choice of energy Lorentz transformation (in turn depending on the unit of analysis): $p' = \gamma^2 p$ or $p' = \gamma^2 (p - ...)$
    \item Looking at the relativistic pressure in terms of its role in the equations of state leads us to conclude that the Lorentz transformation is arbitrary and not unique at all.
    \item Looking at the relativistic pressure in terms of the stress-energy tensor leads us to conclude that the relativistic pressure is both \textit{an}isotropic and Lorentz \textit{non}-invariant. Only the normal stress \textit{in the direction of motion} is transformed: if that direction is the $x$-direction, then as $p'_x = \gamma^2 (p_x + u^2_x \rho)$.
\end{itemize}
Note that these problems go away, and the consilience of these perspectives return, when we return to the situation of $\gamma = 1$, i.e. $\textbf{u} \to 0$. When this is not the case, the pressure's consilience breaks down, just like the temperature, with a myriad of possible Lorentz transformations, contrary to the usual assertion that the Lorentz transformation of pressure \textit{just is} $p' = p$ in the literature surrounding relativistic thermodynamics.

To resist the problem, one might respond that the breakdown of pressure and temperature is simply a result of not taking fundamental physics seriously enough. We've discussed how the pressure concept is defined via four perspectives: via forces, via the fundamental relation, via equations of state, and via continuum mechanics. However, some of these approaches -- the fundamental relation and equations of state in particular -- seem `top-down', in that we are starting with macroscopic, non-fundamental, laws. But why think these laws should be relativistically invariant? We don't expect laws about the regularity of the length of Subway sandwiches to be invariant under Lorentz boosts. Perhaps the `one true' way of defining pressure must appeal to `bottom-up' mechanical approaches, via the force-over-area and continuum mechanical pictures which connects more fundamental physics to pressure (e.g. via mechanical notions like momentum and forces), instead. I have two responses here. Firstly, we could of course ignore the `top-down' approaches and simply \textit{define} whatever we've got from the `bottom-up' approaches as the `one true' pressure concept. However, a worry similar to the case in continuum mechanics (\S3.4) arises: as we've discussed, the `pressure' obtained in that case need not have anything to do with the usual physical concept of pressure in the hydrostatic regime. Likewise, here, without constraint from `top-down' by the fundamental relation between thermodynamic variables (and equations of state), it's not clear that we'd necessarily end up with a concept that has \textit{anything }to do with what we'd ordinarily call pressure outside of the $\textbf{u} \approx 0$ regime. We can, of course, \textit{call} it the `mechanical pressure', but the fact remains that there might not be consilience between the mechanical pressure and the usual thermodynamic pressure. Secondly, \textit{even} if we restricted our attention to the two `bottom-up' approaches, we see that they lead us to \textit{distinct} results: $p' = p$ on one hand, and $p' = \gamma^2 p +...$ on the other. The relativistic pressure concept still falls apart.

\section{Thermodynamic limits and the limits of thermodynamics}

When I discussed $p_{\text{classical}}$ in the context of classical continuum mechanics, I argued that the breakdown of consilience given nontrivial viscosity should not worry us. In the domains where thermodynamics and hydrostatics apply, there exists remarkable consilience for $p_{\text{classical}}$. In the domains where thermodynamics cannot be applied, there is no reason to expect $p_{\text{classical}}$ to apply anyway, and so the breakdown of consilience there should not worry us either. 

I think something similar is happening when it comes to relativizing thermodynamics. The breakdown of $p_{\text{classical}}$ in the relativistic setting seems to signal a limit to the applicability of thermodynamics -- there's no unique way to generalize thermodynamics to render it compatible with the principle of relativity. Each perspective contributed to the meaning of $p_{\text{classical}}$ in a consilient fashion, and the failure to reproduce this consilience in the relativistic setting suggests that the framework no longer holds up. This situation is in no way a fault of thermodynamics. We already think that thermodynamics is non-fundamental when considering how thermodynamics emerges from statistical mechanics. I see my discussion as enriching the sense in which thermodynamics is non-fundamental, by emphasizing that thermodynamics picks out a \textit{preferred frame} in relativity beyond which there is no consilient description of thermodynamic systems: this preferred frame is the rest frame of the system (or system-plus-environment) in equilibrium. Insofar as there is no unique extension of temperature and pressure beyond those regimes, I propose that we view the classical thermodynamic framework as limited only to the rest frame of systems in equilibrium. 

On a standard picture in physics as well as philosophy of physics (e.g. \citet{callender_taking_2001, rau_thermodynamic_2017, batterman_emergence_2011, butterfield_less_2011, palacios_had_2018, wu_explaining_2021}), thermodynamics approximately emerges from statistical mechanics when we take the Thermodynamic Limit: $N \to \infty$, with $\frac{N}{V}$ constant. A lively debate has ensued surrounding the status of this limit, which I won't go into here. I simply note that my discussion suggests \textit{another limit} mandatory for the emergence of classical thermodynamic behavior: for a system in question, the thermodynamic framework only becomes applicable if \textit{it is also the case that} $\textbf{u} \to 0$ for the velocity of the system (or system-plus-environment) relative to the stationary observer (or frame of interest). We've already seen that as the system (or system-plus-environment) moves at a velocity $\textbf{u} \gg 0$ relative to a stationary observer, the concept of classical pressure appears to break down. But if we restricted attention to cases where $\textbf{u} \approx 0$, then we can effectively \textit{just do classical thermodynamics}. Only in this limit do we recover classical thermodynamics in a consilient, unproblematic, and consistent fashion. When $\textbf{u} = 0$, classical thermodynamics is precisely recovered, and all perspectives agree on all the core concepts, including temperature and pressure. In this limit there is no worry about whether any concepts break down, or which generalized concepts to pick as the appropriate `natural' extension of their classical counterparts. When $\textbf{u} \approx 0$, we approximately find overlap between the various concepts, since $\gamma \approx 1$ and all the generalized concepts of pressure (and temperature) approximately agree. However, once $\textbf{u} \gg 0$, the classical concept falls apart, and we can no longer coherently employ the classical thermodynamic framework. 

One sees that the $\textbf{u} \to 0$ limit plays an important and under-emphasized role in ensuring the emergence of thermodynamics from statistical mechanics. A Lorentz-boosted box of particles continues to have a mechanical description, in terms of forces, energy, momentum, positions, and so on, but \textit{it is not guaranteed that it has a unique thermodynamic description}, if we only assumed the standard Thermodynamic Limit $N \to \infty$, and not the $\textbf{u} \to 0$ limit. If both limits are applied, though, a unique classical thermodynamical description arises for this box of particles, i.e. from statistical mechanics. This is what \citet{cubero_thermal_2007} implicitly propose, when they say that any statistical-mechanical derivation of thermodynamic behavior from statistical mechanical equations ``implicitly assumes the presence of a spatial confinement, thus \textit{singling out a preferred frame of reference}.'' Interestingly, \citet[p. 201]{liu_einstein_1992} attributes something similar to Einstein, in a 1953 letter to von Laue, that Einstein implicitly ``makes a special and also fundamental assumption in the equilibrium thermodynamics, namely, \textit{reversible heat exchange can only take place when the bodies involved are at rest with one another}." Such proposals can be made precise with the $\textbf{u} \to 0$ limit, which singles out this preferred frame of reference, a frame in which the relative velocity between system, environment, and the frame itself is zero.\footnote{\citet{ramirez_symmetries_2024} makes a similar claim for ideal springs: Hooke's law requires a rest frame in which the equilibrium point is truly at rest; in other frames the law doesn't apply. In the future I hope to consider the broader physical meaning of this $\textbf{u} \to 0$ limit.}

\section{Concluding remarks}

I've shown that four perspectives in classical physics come together in a consilient fashion to define a unique concept of pressure, $p_{\text{classical}}$ in the classical thermodynamic domain. However, their relativistic counterparts lack this consilience and $p_{\text{classical}}$ breaks down in relativity. When it came to the development of physical concepts across domains, \citet[25]{bridgman1927logic} wrote that ``...a certain haziness is inevitable." Consilience and ``naturalness" are promising tools for finding a unique path through the haze. However, for the case of the relativistic pressure, in line with Einstein, it seems that there is no definitive method for generalizing thermodynamics to relativity. And the situation is a little worse: because of the breakdown of consilience, there is no unique -- ``natural" -- concept of relativistic pressure. 

\section*{Acknowledgments}

I thank Craig Callender, Eddy Keming Chen, Yichen Luo, Wayne Myrvold, Sai Ying Ng, Sebastian Murgueitio Ramirez, Chip Sebens, Shelly Yiran Shi, David Wallace, the audience at UCSD's philosophy of physics reading group, Caltech's brown bag seminar series, the 2024 BSPS annual conference, the foundations of thermodynamics workshop at University of Western Ontario, and the NUS philosophy department's seminar series for their helpful comments and suggestions. I also thank a series of anonymous reviewers for improving the paper. Part of this work was supported by Nanyang Technological University's Nanyang Assistant Professorship Grant for the project ``Philosophical Foundations of Thermodynamics".

\bibliographystyle{plainnat}
\bibliography{references}

@book{Newton1962USP,
  author    = {Isaac Newton},
  editor    = {A. Rupert Hall and Marie Boas Hall},
  title     = {Unpublished Scientific Papers of Isaac Newton},
  publisher = {Cambridge University Press},
  address   = {Cambridge},
  year      = {1962}
}

@article{doi:10.1086/737745,
author = {Chua, Eugene Y. S. and Callender, Craig},
title = {The Toll of the Tolman Effect: On the Status of Classical Temperature in General Relativity},
journal = {The British Journal for the Philosophy of Science},
year = {forthcoming},
doi = {10.1086/737745},

URL = { 
    
        https://doi.org/10.1086/737745
    
    

},
eprint = { 
    
        https://doi.org/10.1086/737745
    
    

}

}

@article{Clausius1857-Heat-PhilMag,
  author  = {Clausius, Rudolf},
  title   = {On the Nature of the Motion which we call Heat},
  journal = {The London, Edinburgh, and Dublin Philosophical Magazine and Journal of Science},
  series  = {4},
  volume  = {14},
  number  = {91},
  pages   = {108--127},
  year    = {1857},
  doi     = {10.1080/14786445708642360},
  note    = {English translation of {\"U}ber die Art der Bewegung, welche wir W{\"a}rme nennen}
}

@article{ramirez_symmetries_2024,
	title = {On {Symmetries} and {Springs}},
	issn = {0031-8248, 1539-767X},
	url = {https://www.cambridge.org/core/journals/philosophy-of-science/article/on-symmetries-and-springs/09DA7C816B63103418D9E5795E9EF582},
	doi = {10.1017/psa.2023.170},
	language = {en},
	urldate = {2024-08-15},
	journal = {Philosophy of Science},
	author = {Ramírez, Sebastián Murgueitio},
	month = jan,
	year = {2024},
	pages = {1--20},
}

@article{chua_t_2023,
	title = {T {Falls} {Apart}: {On} the {Status} of {Classical} {Temperature} in {Relativity}},
	volume = {90},
	issn = {0031-8248, 1539-767X},
	shorttitle = {T {Falls} {Apart}},
	url = {https://www.cambridge.org/core/journals/philosophy-of-science/article/t-falls-apart-on-the-status-of-classical-temperature-in-relativity/43CE3D77A75E3BF5DE46BF37F5671AE1},
	doi = {10.1017/psa.2023.27},
	language = {en},
	number = {5},
	urldate = {2023-12-20},
	journal = {Philosophy of Science},
	author = {Chua, Eugene Yew Siang},
	month = dec,
	year = {2023},
	note = {Publisher: Cambridge University Press},
	pages = {1307--1319},
}

@article{uffink_bluff_2001,
	title = {Bluff {Your} {Way} in the {Second} {Law} of {Thermodynamics}},
	volume = {32},
	doi = {10.1016/s1355-2198(01)00016-8},
	number = {3},
	journal = {Studies in History and Philosophy of Science Part B: Studies in History and Philosophy of Modern Physics},
	author = {Uffink, Jos},
	year = {2001},
	pages = {305--394},
}

@article{landsberg_a_1967,
	title = {À relativistic generalization of thermodynamics},
	volume = {52},
	issn = {1826-9877},
	url = {https://doi.org/10.1007/BF02710651},
	doi = {10.1007/BF02710651},
	language = {en},
	number = {1},
	urldate = {2023-11-01},
	journal = {Il Nuovo Cimento B (1965-1970)},
	author = {Landsberg, P. and Johns, K.},
	month = nov,
	year = {1967},
	pages = {28--44},
}

@book{mihalas_foundations_1984,
	title = {Foundations of radiation hydrodynamics},
	url = {https://ui.adsabs.harvard.edu/abs/1984oup..book.....M},
	urldate = {2023-12-09},
	author = {Mihalas, D. and Mihalas, B. W.},
	month = jan,
	year = {1984},
}

@article{cubero_thermal_2007,
	title = {Thermal equilibrium and statistical thermometers in special relativity},
	volume = {99},
	issn = {0031-9007, 1079-7114},
	url = {http://arxiv.org/abs/0705.3328},
	doi = {10.1103/PhysRevLett.99.170601},
	number = {17},
	urldate = {2023-12-07},
	journal = {Physical Review Letters},
	author = {Cubero, David and Casado-Pascual, Jesús and Dunkel, Jörn and Talkner, Peter and Hänggi, Peter},
	month = oct,
	year = {2007},
	note = {arXiv:0705.3328 [astro-ph, physics:cond-mat, physics:hep-th]},
	pages = {170601},
}

@article{wu_explaining_2021,
	title = {Explaining universality: infinite limit systems in the renormalization group method},
	volume = {199},
	issn = {1573-0964},
	shorttitle = {Explaining universality},
	url = {https://doi.org/10.1007/s11229-021-03448-2},
	doi = {10.1007/s11229-021-03448-2},
	language = {en},
	number = {5},
	urldate = {2023-12-07},
	journal = {Synthese},
	author = {Wu, Jingyi},
	month = dec,
	year = {2021},
	pages = {14897--14930},
}

@article{callender_taking_2001,
	title = {Taking {Thermodynamics} {Too} {Seriously}},
	volume = {32},
	doi = {10.1016/s1355-2198(01)00025-9},
	number = {4},
	journal = {Studies in History and Philosophy of Science Part B: Studies in History and Philosophy of Modern Physics},
	author = {Callender, Craig},
	year = {2001},
	pages = {539--553},
}

@article{butterfield_less_2011,
	title = {Less is {Different}: {Emergence} and {Reduction} {Reconciled}},
	volume = {41},
	issn = {1572-9516},
	shorttitle = {Less is {Different}},
	url = {https://doi.org/10.1007/s10701-010-9516-1},
	doi = {10.1007/s10701-010-9516-1},
	language = {en},
	number = {6},
	urldate = {2023-12-07},
	journal = {Foundations of Physics},
	author = {Butterfield, J.},
	month = jun,
	year = {2011},
	pages = {1065--1135},
}

@article{batterman_emergence_2011,
	title = {Emergence, {Singularities}, and {Symmetry} {Breaking}},
	volume = {41},
	issn = {1572-9516},
	url = {https://doi.org/10.1007/s10701-010-9493-4},
	doi = {10.1007/s10701-010-9493-4},
	language = {en},
	number = {6},
	urldate = {2023-12-07},
	journal = {Foundations of Physics},
	author = {Batterman, Robert W.},
	month = jun,
	year = {2011},
	pages = {1031--1050},
}

@article{palacios_had_2018,
	title = {Had {We} {But} {World} {Enough}, and {Time}... {But} {We} {Don}’t!: {Justifying} the {Thermodynamic} and {Infinite}-{Time} {Limits} in {Statistical} {Mechanics}},
	volume = {48},
	issn = {1572-9516},
	shorttitle = {Had {We} {But} {World} {Enough}, and {Time}... {But} {We} {Don}’t!},
	url = {https://doi.org/10.1007/s10701-018-0165-0},
	doi = {10.1007/s10701-018-0165-0},
	language = {en},
	number = {5},
	urldate = {2023-12-07},
	journal = {Foundations of Physics},
	author = {Palacios, Patricia},
	month = may,
	year = {2018},
	pages = {526--541},
}

@incollection{rau_thermodynamic_2017,
	title = {Thermodynamic {Limit}},
	urldate = {2023-12-07},
	booktitle = {Statistical {Physics} and {Thermodynamics}: {An} {Introduction} to {Key} {Concepts}},
	publisher = {Oxford University Press},
	author = {Rau, Jochen},
	editor = {Rau, Jochen},
	month = sep,
	year = {2017},
	doi = {10.1093/oso/9780199595068.003.0005},
}

@article{penney_note_1966,
	title = {Note on relativistic thermodynamics},
	volume = {43},
	issn = {1826-9869},
	url = {https://doi.org/10.1007/BF02756369},
	doi = {10.1007/BF02756369},
	language = {en},
	number = {4},
	urldate = {2023-12-03},
	journal = {Il Nuovo Cimento A (1965-1970)},
	author = {Penney, R.},
	month = jun,
	year = {1966},
	pages = {911--918},
}

@article{kibble_relativistic_1966,
	title = {Relativistic transformation laws for thermodynamic variables},
	volume = {41},
	issn = {1826-9877},
	url = {https://doi.org/10.1007/BF02711119},
	doi = {10.1007/BF02711119},
	language = {en},
	number = {1},
	urldate = {2023-12-03},
	journal = {Il Nuovo Cimento B (1965-1970)},
	author = {Kibble, T. W. B.},
	month = jan,
	year = {1966},
	pages = {72--78},
}

@article{ott_lorentz-transformation_1963,
	title = {Lorentz-{Transformation} der {Wärme} und der {Temperatur}},
	volume = {175},
	issn = {0044-3328},
	url = {https://doi.org/10.1007/BF01375397},
	doi = {10.1007/BF01375397},
	language = {de},
	number = {1},
	urldate = {2023-12-03},
	journal = {Zeitschrift für Physik},
	author = {Ott, H.},
	month = feb,
	year = {1963},
	pages = {70--104},
}

@article{arzelies_transformation_1965,
	title = {Transformation relativiste de la température et de quelques autres grandeurs thermodynamiques},
	volume = {35},
	issn = {1827-6121},
	url = {https://doi.org/10.1007/BF02739342},
	doi = {10.1007/BF02739342},
	language = {fr},
	number = {3},
	urldate = {2023-12-03},
	journal = {Il Nuovo Cimento (1955-1965)},
	author = {Arzeliès, H.},
	month = feb,
	year = {1965},
	pages = {792--804},
}

@article{gamba_relativistic_1965,
	title = {Relativistic transformarion of thermodynamical quantities. ({Beware} of {Jacobians}!)},
	volume = {37},
	issn = {1827-6121},
	url = {https://doi.org/10.1007/BF02783385},
	doi = {10.1007/BF02783385},
	language = {en},
	number = {4},
	urldate = {2023-12-03},
	journal = {Il Nuovo Cimento (1955-1965)},
	author = {Gamba, A.},
	month = jun,
	year = {1965},
	pages = {1792--1794},
}

@book{reichl_modern_1980,
	title = {A {Modern} {Course} in {Statistical} {Physics}},
	isbn = {978-0-7131-2777-5},
	language = {en},
	publisher = {E. Arnold},
	author = {Reichl, L. E.},
	year = {1980},
}

@book{callen_thermodynamics_1991,
	title = {Thermodynamics and an {Introduction} to {Thermostatistics}},
	isbn = {978-0-471-86256-7},
	language = {en},
	publisher = {John Wiley \& Sons},
	author = {Callen, Herbert B.},
	month = jan,
	year = {1991},
}

@book{truesdell_first_1992,
	title = {A {First} {Course} in {Rational} {Continuum} {Mechanics} {V1}},
	isbn = {978-0-08-087387-9},
	language = {en},
	publisher = {Academic Press},
	author = {Truesdell, C. A.},
	month = feb,
	year = {1992},
}

@book{tadmor_continuum_2012,
	title = {Continuum {Mechanics} and {Thermodynamics}: {From} {Fundamental} {Concepts} to {Governing} {Equations}},
	isbn = {978-1-107-00826-7},
	shorttitle = {Continuum {Mechanics} and {Thermodynamics}},
	language = {en},
	publisher = {Cambridge University Press},
	author = {Tadmor, Ellad B. and Miller, Ronald E. and Elliott, Ryan S.},
	year = {2012},
}

@book{irgens_continuum_2008,
	title = {Continuum {Mechanics}},
	isbn = {978-3-540-74298-2},
	language = {en},
	publisher = {Springer Science \& Business Media},
	author = {Irgens, Fridtjov},
	month = jan,
	year = {2008},
}

@book{birkhoff_hydrodynamics_2015,
	title = {Hydrodynamics},
	isbn = {978-1-4008-7777-5},
	language = {en},
	publisher = {Princeton University Press},
	author = {Birkhoff, Garrett},
	month = dec,
	year = {2015},
}

@misc{brechet_relativistic_2022,
	title = {Relativistic thermodynamics of perfect fluids},
	url = {http://arxiv.org/abs/2210.04282},
	doi = {10.48550/arXiv.2210.04282},
	urldate = {2023-11-29},
	publisher = {arXiv},
	author = {Brechet, Sylvain D. and Girard, Marin C. A.},
	month = oct,
	year = {2022},
	note = {arXiv:2210.04282 [cond-mat, physics:physics]},
}

@book{white_viscous_1991,
	edition = {2nd},
	title = {Viscous fluid flow},
	publisher = {McGraw-Hill New York},
	author = {White, Frank M},
	year = {1991},
}

@article{cauchy_pression_1827,
	title = {De la pression ou tension dans un corps solide, [{On} the pressure or tension in a solid body},
	volume = {2},
	language = {fr},
	journal = {Exercices de Mathématiques},
	author = {{Cauchy}},
	year = {1827},
	pages = {42},
}

@book{anderson_equations_1995,
	address = {Oxford, New York},
	series = {Oxford {Monographs} on {Geology} and {Geophysics}},
	title = {Equations of {State} of {Solids} in {Geophysics} and {Ceramic} {Science}},
	isbn = {978-0-19-505606-8},
	publisher = {Oxford University Press},
	author = {Anderson, Orson L.},
	month = apr,
	year = {1995},
}

@book{halliday_fundamentals_2007,
	edition = {8},
	title = {Fundamentals of physics, extended edition},
	publisher = {Wiley},
	author = {Halliday, David and Resnick, Robert and Walker, Jearl},
	year = {2007},
}

@article{euler_principes_1757,
	title = {Principes généraux du mouvement des fluides},
	url = {https://scholarlycommons.pacific.edu/euler-works/226},
	journal = {Mémoires de l'académie des sciences de Berlin},
	author = {Euler, Leonhard},
	month = jan,
	year = {1757},
	pages = {274--315},
}

@book{newton_principia_1687,
	title = {The {Principia}: {The} {Authoritative} {Translation} and {Guide}: {Mathematical} {Principles} of {Natural} {Philosophy}},
	language = {en},
	urldate = {2023-11-22},
	author = {Newton, Isaac},
	translator = {Cohen, I. Bernard and Whitman, Anne and Budenz, Julia},
	year = {1687},
}

@book{massey_mechanics_2006,
	address = {London},
	edition = {8},
	title = {Mechanics of {Fluids}},
	publisher = {Taylor and Francis},
	author = {Massey, Bernard and Ward-Smith, John},
	year = {2006},
}

@article{chalmers_how_2018,
	title = {How {Pressure} {Became} a {Scalar}, {Not} a {Vector}},
	volume = {20},
	issn = {1422-6960},
	url = {https://doi.org/10.1007/s00016-018-0221-3},
	doi = {10.1007/s00016-018-0221-3},
	language = {en},
	number = {2},
	urldate = {2023-11-22},
	journal = {Physics in Perspective},
	author = {Chalmers, A.},
	month = jun,
	year = {2018},
	pages = {165--179},
}

@book{chalmers_one_2017,
	address = {Cham},
	series = {Archimedes},
	title = {One {Hundred} {Years} of {Pressure}},
	volume = {51},
	url = {http://link.springer.com/10.1007/978-3-319-56529-3},
	urldate = {2023-11-22},
	publisher = {Springer International Publishing},
	author = {Chalmers, A.},
	year = {2017},
	doi = {10.1007/978-3-319-56529-3},
}

@article{balescu_relativistic_1968,
	title = {Relativistic statistical thermodynamics},
	volume = {40},
	issn = {0031-8914},
	url = {https://www.sciencedirect.com/science/article/pii/0031891468901328},
	doi = {10.1016/0031-8914(68)90132-8},
	number = {3},
	urldate = {2023-11-01},
	journal = {Physica},
	author = {Balescu, R.},
	month = dec,
	year = {1968},
	pages = {309--338},
}

@incollection{loewer_mentaculus_2019,
	title = {The {Mentaculus} {Vision}},
	isbn = {9789811211713},
	url = {https://www.worldscientific.com/doi/10.1142/9789811211720_0001},
	urldate = {2023-10-30},
	booktitle = {Statistical {Mechanics} and {Scientific} {Explanation}},
	publisher = {World Scientific},
	author = {Loewer, Barry},
	month = aug,
	year = {2019},
	doi = {10.1142/9789811211720_0001},
	pages = {3--29},
}

@book{pauli_theory_1958,
	address = {New York,},
	title = {Theory of {Relativity}},
	publisher = {Pergamon Press},
	author = {Pauli, Wolfgang},
	year = {1958},
}

@book{tolman_relativity_1934,
	address = {Oxford,},
	title = {Relativity, {Thermodynamics} and {Cosmology}},
	publisher = {Clarendon Press},
	author = {Tolman, Richard Chace},
	year = {1934},
}

@article{sutcliffe_lorentz_1965,
	title = {Lorentz transformations of thermodynamic quantities},
	volume = {39},
	issn = {1827-6121},
	url = {https://doi.org/10.1007/BF02735833},
	doi = {10.1007/BF02735833},
	language = {en},
	number = {2},
	urldate = {2023-11-01},
	journal = {Il Nuovo Cimento (1955-1965)},
	author = {Sutcliffe, W. G.},
	month = sep,
	year = {1965},
	pages = {683--686},
}

@article{farias_what_2017,
	title = {What is the temperature of a moving body?},
	volume = {7},
	copyright = {2017 The Author(s)},
	issn = {2045-2322},
	url = {https://www.nature.com/articles/s41598-017-17526-4},
	doi = {10.1038/s41598-017-17526-4},
	language = {en},
	number = {1},
	urldate = {2023-11-01},
	journal = {Scientific Reports},
	author = {Farías, Cristian and Pinto, Victor A. and Moya, Pablo S.},
	month = dec,
	year = {2017},
	note = {Number: 1
Publisher: Nature Publishing Group},
	pages = {17657},
}

@article{liu_einstein_1992,
	title = {Einstein and {Relativistic} {Thermodynamics} in 1952: {A} {Historical} and {Critical} {Study} of a {Strange} {Episode} in the {History} of {Modern} {Physics}},
	volume = {25},
	shorttitle = {Einstein and {Relativistic} {Thermodynamics} in 1952},
	doi = {10.1017/s0007087400028764},
	number = {2},
	journal = {British Journal for the History of Science},
	author = {Liu, Chuang},
	year = {1992},
	note = {Publisher: Cambridge University Press},
	pages = {185--206},
}

@article{liu_is_1994,
	title = {Is {There} a {Relativistic} {Thermodynamics}? {A} {Case} {Study} of the {Meaning} of {Special} {Relativity}},
	volume = {25},
	shorttitle = {Is {There} a {Relativistic} {Thermodynamics}?},
	doi = {10.1016/0039-3681(94)90073-6},
	number = {6},
	journal = {Studies in History and Philosophy of Science Part A},
	author = {Liu, Chuang},
	year = {1994},
	note = {Publisher: Elsevier},
	pages = {983--1004},
}

@book{atkins_four_2007,
	address = {Oxford, New York},
	title = {Four {Laws} {That} {Drive} the {Universe}},
	isbn = {978-0-19-923236-9},
	publisher = {Oxford University Press},
	author = {Atkins, Peter},
	month = sep,
	year = {2007},
}

@article{einstein_relativity_1907,
	title = {On the relativity principle and the conclusions drawn from it},
	volume = {4},
	url = {https://cir.nii.ac.jp/crid/1572543024916682240},
	urldate = {2023-10-21},
	journal = {Jahrb Radioaktivitat Elektronik},
	author = {Einstein, Albert},
	year = {1907},
	pages = {411--462},
}

@article{planck_dynamics_1908,
	title = {On the {Dynamics} of {Moving} {Systems}},
	volume = {331},
	number = {6},
	journal = {Annalen der Physik},
	author = {Planck, Max},
	year = {1908},
	pages = {1--34},
}

@misc{dougherty_black_2016,
	type = {Preprint},
	title = {Black {Hole} {Thermodynamics}: {More} {Than} an {Analogy}?},
	shorttitle = {Black {Hole} {Thermodynamics}},
	url = {https://philsci-archive.pitt.edu/13195/},
	language = {en},
	urldate = {2023-10-21},
	author = {Dougherty, John and Callender, Craig},
	month = oct,
	year = {2016},
}

@book{bridgman1927logic,
  author    = {Bridgman, P. W.},
  title     = {The Logic of Modern Physics},
  year      = {1927},
  publisher = {Macmillan Co.},
  address   = {New York}
}

@book{weyl1949philosophy,
  author    = {Hermann Weyl},
  title     = {Philosophy of Mathematics and Natural Science},
  year      = {1949},
  publisher = {Princeton University Press},
  address   = {Princeton}
}

@article{wallace_case_2018,
	title = {The {Case} for {Black} {Hole} {Thermodynamics} {Part} {I}: {Phenomenological} {Thermodynamics}},
	volume = {64},
	shorttitle = {The {Case} for {Black} {Hole} {Thermodynamics} {Part} {I}},
	doi = {10.1016/j.shpsb.2018.05.002},
	journal = {Studies in History and Philosophy of Science Part B: Studies in History and Philosophy of Modern Physics},
	author = {Wallace, David},
	year = {2018},
	pages = {52--67},
}

@article{chen_quantum_2021,
	title = {Quantum {Mechanics} in a {Time}-{Asymmetric} {Universe}: {On} the {Nature} of the {Initial} {Quantum} {State}},
	volume = {72},
	shorttitle = {Quantum {Mechanics} in a {Time}-{Asymmetric} {Universe}},
	doi = {10.1093/bjps/axy068},
	number = {4},
	journal = {British Journal for the Philosophy of Science},
	author = {Chen, Eddy Keming},
	year = {2021},
	note = {Publisher: University of Chicago Press},
	pages = {1155--1183},
}

@book{albert_time_2000,
	address = {Cambridge, Mass.},
	title = {Time and {Chance}},
	publisher = {Harvard University Press},
	author = {Albert, David Z.},
	year = {2000},
}

\end{document}